\documentclass[11pt,twoside]{article}
\usepackage[T1]{fontenc}
\usepackage{amssymb,amsmath,amsfonts,dsfont,amsthm,bm}
\usepackage{float,rotfloat} 
\usepackage{mathrsfs}
\usepackage{multirow}
\usepackage{enumerate}
\usepackage{enumitem}
\usepackage{mathtools}
\usepackage{hyperref,url}
\usepackage{graphicx}
\usepackage[hang,small,bf]{caption}
\usepackage{subcaption}
\usepackage{booktabs}
\usepackage[top=1in, bottom=1in, left=1.0in, right=1.0 in]{geometry}
\usepackage{longtable}
\usepackage{color,xcolor}
\usepackage[round,sort,comma]{natbib}

\usepackage{graphicx} 
\usepackage{epsfig}
\usepackage[title]{appendix}
\usepackage{etoolbox} 
\usepackage[english]{babel}
\usepackage[autostyle, english = american]{csquotes}
\usepackage{orcidlink}
\MakeOuterQuote{"}
\usepackage{mathtools}
\usepackage{fancyhdr}

\patchcmd{\appendices}{\quad}{: }{}{} 

\newtheorem{thm}{Theorem}

\newtheorem{lemma}{Lemma} 
\newtheorem{cor}{Corollary}

\newtheorem{note}{Note}

\newcommand{\ol}{\overline}
\newcommand{\E}{\mathbb{E}}
\newcommand{\Var}{\mathbb{V}ar}
\newcommand{\R}{\mathbb{R}}

\definecolor{darkblue}{rgb}{0,0,0.55}

\newcommand{\blue}{\color[rgb]{0,0,1}}

\definecolor{dkgreen}{rgb}{0,0.6,0}

\allowdisplaybreaks

\hypersetup{,colorlinks=true,allcolors=darkblue}

\begin{document}   
	
	\baselineskip 5mm
	
	\thispagestyle{empty}
	
	\begin{center}
		
		{\LARGE\bf
			$L$-estimation of Claim Severity Models \\[10pt]
			Weighted by Kumaraswamy Density
		}
		
		\vspace{5mm}
		
		{\large\sc
			Chudamani Poudyal\footnote[1]
			{
				{\sc Corresponding Author}:
				Chudamani Poudyal, PhD, ASA, 
				is an Assistant Professor 
				in the Department of Statistics and Data Science,
				University of Central Florida, 
				Orlando, FL 32816, USA. 
				~~ {\em e-mail\/}: ~{\tt Chudamani.Poudyal@ucf.edu}}}%
		\orcidlink{0000-0003-4528-867X}
		
		\vspace{1mm}
		
		{\large\em University of Central Florida}
		
		\vspace{4mm}
		
		{\large\sc
			Gokarna R. Aryal\footnote[2]{
				Gokarna R. Aryal, Ph.D., is a Professor 
				in the Department of 
				Mathematics, Statistics \& CS,
				Purdue University Northwest, 
				Hammond, IN 46323, USA. 
				{\em e-mail\/}: 
				~{\tt aryalg@pnw.edu}}}
		
		\vspace{1mm}
		
		{\large\em Purdue University Northwest}
		
		\vspace{4mm}
		
		{\large\sc
			Keshav Pokhrel\footnote[3]{
				Keshav Pokhrel, Ph.D., is an Associate Professor 
				in the Department of Mathematics and Statistics,
				University of Michigan-Dearborn, 
				Dearborn, MI 48128, USA. 
				{\em e-mail\/}: 
				~{\tt kpokhrel@umich.edu}}}
		
		\vspace{1mm}
		
		{\large\em University of Michigan-Dearborn}
		
		\vspace{5mm}
		
		\copyright \
		Copyright of this Manuscript is held by the Authors! 
		
	\end{center}
	
	\vspace{0mm}
	
	\begin{quote}
		{\bf\em Abstract:\/}
		Statistical modeling of claim severity distributions 
		is essential in insurance and risk management, 
		where achieving a balance between robustness 
		and efficiency in parameter estimation
		is critical against model contaminations. 
		Two \( L \)-estimators, 
		the method of trimmed moments (MTM) and 
		the method of winsorized moments (MWM), 
		are commonly used in the literature,
		but they are constrained by rigid weighting schemes 
		that either discard or uniformly down-weight extreme observations, 
		limiting their customized adaptability. 
		This paper proposes a flexible robust \( L \)-estimation
		framework weighted by Kumaraswamy densities, 
		offering smoothly varying observation-specific weights
		that preserve valuable information while improving robustness 
		and efficiency. 
		The framework is developed for parametric claim severity models, 
		including Pareto, lognormal, and Fr{\'e}chet distributions, 
		with theoretical justifications on asymptotic normality 
		and variance-covariance structures. 
		Through simulations and application to a U.S. indemnity loss dataset, 
		the proposed method demonstrates superior performance over MTM, MWM, 
		and MLE approaches, 
		particularly in handling outliers and heavy-tailed distributions, 
		making it a flexible and reliable alternative for loss severity modeling.
		
		\vspace{3mm}
		
		\vspace{4mm}
		
	\end{quote}
	
	\newpage
	
	\baselineskip 7mm
	\thispagestyle{empty}
	\tableofcontents
	
	\newpage
	
	
	\baselineskip 7mm
	\setcounter{page}{1}
	
	\section{Introduction}
	\label{sec:Intro}
	
	The modeling and estimation of claim severity distributions are fundamental challenges in insurance and risk management, significantly impacting the premium pricing, reserve determination, and risk assessment. The accuracy and robustness of actuarial calculations are heavily influences by the choice of parameter estimation techniques. 
	Traditional methods, such as maximum likelihood estimation (MLE), often struggles to give accurate estimations due to the presence of outliers and heavy-tailed distributions, prompting the need of alternative methods. To address the need for robustness and the development of less sensitive fitted models, numerous
	research efforts have focused on mitigating the impact of outlier contamination. Practically all of them can be found as special cases of some general classes of statistics, such as $M$-, $L$-, and $R$-statistics \cite{MR2497558}. 
	The robust $M$-estimators, including MLE, have been extensively studied for generalized linear models, as discussed in \cite{MR3132477} and the references therein. In the context of actuarial loss modeling, \cite{MR4477474} introduced the maximum weighted likelihood estimator (MWLE) for robust tail estimation within finite mixture models. Building on this foundation, \cite{tcf24} proposed score-based weighted likelihood estimation (SWLE), specifically designed for robust estimation in generalized linear models (GLMs). 
	
	The $L$-estimators which are linear combinations of the functions of order statistics have gained lots of attention for their resilience, offering a robust framework for actuarial and financial applications. Two widely used robust $L$-estimators in loss modeling for fully observed ground-up loss data are the \emph{method of trimmed moments} (MTM)  \citep[see, e.g.,][]{MR2497558} and the \emph{method of winsorized moments} (MWM) \citep[see, e.g.,][]{MR3758788}. The MTM disregards a fixed proportion of observations corresponding to the trimming proportions, effectively discarding the information from these trimmed sample values, which are deemed outliers. 
	In contrast, MWM retains all observed values and assigns reduced weights to extreme sample observations, thereby enhancing robustness. Nevertheless, both MTM and MWM apply uniform weights to the remaining observations, potentially overlooking subtle variations in the data distribution. 
	A series of works,
	including 
	\cite{MR4263275},  
	\cite{pb22},
	\cite{MR4602526}, and 
	\cite{MR4712558}, 
	have extended the application of MTM and MWM estimators to scenarios involving incomplete, truncated, or censored data. 
	These studies demonstrate that trimming and winsorizing are effective approaches for enhancing the robustness of moment estimation in the presence of extreme claims, 
	particularly by mitigating the impact of heavier point masses at the left 
	truncation and right censoring points \cite{gw24}.
	
	Although both MTM and MWM methods aim to improve robustness in the presence of outliers and heavy-tailed distributions, but they come with trade-offs in terms of data loss, potential bias, and the need for careful selection of parameters.   Motivated by these limitations of MTM and MWM, this study proposes a flexible robust $L$-estimation framework that generates smoothly varying weights across the entire data range, enabling more nuanced and adaptive modeling. Unlike MTM, which discards extreme observations, and MWM, which applies uniform down-weighting to them, the proposed methodology employs observation-specific weights to preserve and incorporate valuable information from the full dataset. In particular, we propose to enhance $L$-estimation framework by incorporating weights derived from Kumaraswamy density functions, offering a flexible and observation-specific weighting mechanism that improves the robustness and adaptability of these estimators. 
	
	The remainder of the paper is organized as follows. Section \ref{sec:Methodology} provides a brief overview of the robust methodology for $L$-estimators weighted by the Kumaraswamy density. In Section \ref{sec:ParametricModels}, the theoretical framework is developed, presenting explicit formulations of the proposed $L$-estimators for various parametric claim severity models. This section also includes derivations of the asymptotic normality and variance-covariance structure of the estimators. Section \ref{sec:SimStudy} offers a comprehensive simulation study to validate the theoretical findings and assess the finite-sample performance of the estimators. In Section \ref{sec:RealDataAnalysis}, the proposed methodology is applied to real-world data, with its performance thoroughly analyzed. Finally, Section \ref{sec:Conclusion} provides concluding remarks and suggests directions for future research.
	
	
	\section{Methodology}
	\label{sec:Methodology}
	
	This section is divided into two parts.
	The first part provides a summary of the structural development 
	of \( L \)-estimators along with their inferential justification. 
	The second part investigates the Kumaraswamy weighting mechanism
	in detail, aiming to achieve a desired balance between robustness
	and efficiency in \( L \)-estimators.
	
	\subsection{Robust {\em L} -- Estimators}
	\label{sec:L_Estimator}
	
	For a positive integer $n$,
	let $X_{1}, \ldots, X_{n}$ be an iid sample
	from an unknown true underlying cumulative 
	distribution function $F$ with the parameter vector 
	$\bm{\theta} = (\theta_{1}, \cdots, \theta_{k})$.
	The motivation of most of the statistical inference 
	is to estimate the parameter vector $\bm{\theta}$
	from the available sample dataset.
	The corresponding order statistic
	of the sample is denoted by 
	$X_{1:n}, \ldots, X_{n:n}$.
	In order to estimate the 
	parameter vector $\bm{\theta}$,
	the statistics we are interested 
	here is a {\em linear combination}
	of the order values,
	so the name $L$-statistics, 
	in the form 
	\begin{align}
		\label{eqn:S1}
		\widehat{\mu}_{j} 
		& :=
		\dfrac{1}{n}
		\sum_{i=1}^{n}
		J_{j} \left( \dfrac{i}{n+1} \right) 
		h_{j}(X_{i:n}),
		\quad 
		1 \le j \le k,
	\end{align}
	where $J_{j} : [0,1] \to \R_{\ge 0}$
	represents a {\em weights-generating} function.
	Both $J_{j}$ and $h_{j}$
	are specially chosen functions,
	\citep[see, e.g.,][]{MR4263275, pb22}
	and are known that are specified
	by the statistician.
	
	The corresponding population quantities
	are then given by 
	\begin{align}
		\label{eqn:P1}
		\mu_{j}
		& \equiv 
		\mu_{j} 
		\left( 
		\bm{\theta}
		\right) 
		\equiv 
		\mu_{j} 
		\left( 
		\theta_1, \ldots, \theta_k
		\right) 
		= 
		\int_{0}^{1}
		{J_{j}(u)H_{j}(u) \, du,
			\quad 
			1\leq{j}\leq{k}},
		\mbox{ where } 
		H_{j} := h_{j} \circ F^{-1}.
	\end{align}
	
	$L$-estimators \citep{MR1049304} are found by matching sample 
	$L$-moments,
	Eq. \eqref{eqn:S1}, 
	with population $L$-moments, 
	Eq. \eqref{eqn:P1},
	for $j = 1, \ldots, k$,
	and then solving the system of equations
	with respect to $\theta_1, \ldots, \theta_k$.
	The obtained solutions, 
	which we denote by 
	$
	\widehat{\theta}_j 
	=
	g_j(\widehat{\mu}_{1}, \ldots, \widehat{\mu}_{k})$,
	$1 \leq j \leq k$, are, by definition, 
	the $L$-estimators of 
	$\theta_1, \ldots, \theta_k$. 
	Note that the functions $g_j$ are such that
	$
	\theta_j 
	=
	g_j(\mu_1(\boldsymbol{\theta}),
	\ldots, \mu_k(\boldsymbol{\theta})).
	$
	
	Define $\ol{u} = 1-u$, 
	and consider 
	\begin{align}
		\label{eqn:AlphaFun1}
		\alpha_{j}(u)
		& = 
		\frac{1}{\ol{u}}
		\int_{u}^{1}
		\ol{v} J_{j}(v)H_{j}'(v) \, dv, 
		\quad 
		1\leq{j}\leq{k}.
	\end{align}
	
	Further, let
	\[
	\widehat{\bm\mu}
	:=
	\left( 
	\widehat{\mu}_{1}, 
	\widehat{\mu}_{1}, 
	\ldots, 
	\widehat{\mu}_{k}
	\right)
	\quad 
	\mbox{and}
	\quad 
	\bm{\mu}
	:=
	\left(
	\mu_{1}, 
	\mu_{2}, 
	\ldots, 
	\mu_{k}
	\right).
	\]
	
	Ideally, 
	we expect that the statistics vector 
	$\widehat{\bm\mu}$
	converges in distribution 
	to the population vector $\bm{\mu}$.
	As mentioned by 
	\citet[][\S8.2 and references therein]{MR595165}, 
	there are several approaches of 
	establishing asymptotic normality
	of $\widehat{\bm\mu}$ 
	depending upon the various 
	scenarios of the weights generating
	function $J$ and the underlying cdf $F$.
	
	\begin{thm}[\citealp{MR0203874}, {\sc Remark 9}]
		\label{thm:CGJ1}
		The $k$-variate vector
		$\sqrt{n}(\widehat{\bm\mu}-\bm{\mu})$,
		converges in distribution to the 
		$k$-variate normal random vector
		with mean $\mathbf{0}$ and 
		the variance-covariance matrix $\mathbf{\Sigma}:= \left[\sigma_{ij}^{2}\right]_{i,j=1}^{k}$ 
		with the entries 
		\begin{align}
			\sigma_{ij}^{2} 
			& = 
			\int_{0}^{1} 
			{\alpha_{i}(u)\alpha_{j}(u) \, du}  
			= 
			\int_{0}^{1} 
			\int_{0}^{1} 
			J_{i}(v) J_{j}(w)
			K(v,w) \, 
			dH_{i}(v) \, 
			dH_{j}(w),
			\label{eqn:mtm_var_cov3}
		\end{align}
		where the function $K(v,w)$ is defined as
		\begin{align}
			\label{eqn:kFun1}
			K(v,w) 
			& := 
			K(w, v)
			=
			v \wedge w 
			- 
			vw
			= 
			\min \{v, w \} - vw,
			\quad 
			\mbox{for} 
			\quad 
			0 \le v, w \le 1.
		\end{align}
	\end{thm}
	
	Now, with 
	$\bm{\widehat{\mu}}
	=
	\left(\widehat{\mu}_{1},\ldots,\widehat{\mu}_{k} \right)$ 
	and
	$\theta_j 
	=
	g_j(\mu_1(\boldsymbol{\theta}),
	\ldots, \mu_k(\boldsymbol{\theta})),
	$
	then by delta method \citep[see, e.g.,][Theorem A, p. 122]{MR595165}, 
	we state the following asymptotic result.
	
	\begin{thm} 
		\label{thm:CGJ2}
		The $L$-estimator of $\bm{\theta}$, 
		denoted by
		$\widehat{\bm{\theta}}$, 
		has the following asymptotic distribution: 
		\begin{eqnarray}
			\label{eqn:DeltaMethod1}
			\widehat{\bm{\theta}} 
			& = &
			\left(\widehat{\theta}_{1},\ldots,\widehat{\theta}_{k}\right) 
			\sim \mathcal{AN}\left(\bm{\theta},\frac{1}{n}{\bm{D\Sigma D'}}\right),
		\end{eqnarray}
		where the Jacobian $\bm{D}$ is given by
		$
		\bm{D}
		=
		\left[\left. \frac{\partial g_{i}}{\partial \widehat{\mu}_{j}}\right\vert_{\widehat{\bm{\mu}}
			=
			\bm{\mu}}\right]_{k\times k} 
		=:
		\left[d_{ij}\right]_{k\times k}
		$ and the variance-covariance matrix $\bm{\Sigma}$ has 
		the same form as in Theorem \ref{thm:CGJ1}.
	\end{thm}
	
	For specific choices of the weight-generating functions \( J_{j}, \, 1 \le j \le k \), as defined in Eq.~\eqref{eqn:S1}, the resulting estimators reduce to MTM or MWM. Further details regarding these approaches can be found in \cite{cp24}. 
	We conclude this section with the following result,
	which will be used in subsequent discussions.
	
	\begin{thm}
		\label{thm:GeneralCSIneq1}
		Let \( f_{i} : (0,1) \to \R \), \( 1 \leq i \leq 2 \), 
		be two non-zero functions such that 
		\( f_{1}, f_{2} \in L^{2} \)-space
		and \( f_{1} \) and \( f_{2} \) 
		are not linearly dependent. 
		Consider the following integrals:
		\begin{eqnarray*}
			\Omega_{1} 
			& := & 
			\int_{0}^{1} \int_{0}^{1} 
			f_{1}(x) \, f_{1}(y) \, K(x, y) \, dy \, dx, \\
			\Omega_{2} 
			& := & 
			\int_{0}^{1} \int_{0}^{1} 
			f_{1}(x) \, f_{1}(y) \, f_{2}(y) \, K(x, y) \, dy \, dx, \\
			\Omega_{3} 
			& := & 
			\int_{0}^{1} \int_{0}^{1} 
			f_{1}(x) \, f_{1}(y) \, f_{2}(x) \, f_{2}(y)\, K(x, y) \, dy \, dx, 
		\end{eqnarray*}
		where $K(\cdot, \cdot)$ is defined in 
		Eq.~\eqref{eqn:kFun1}.
		Then, the following inequality holds: 
		\begin{eqnarray*}
			\Omega_{2}^{2} 
			& < & 
			\Omega_{1} \, \Omega_{3}.
		\end{eqnarray*}
		
		\begin{proof}
			From \citet[][Section 4.7]{MR2057928}, the kernel function \( K(x, y) \) represents the covariance of the Brownian bridge \( B(t) \) for \( t \in [0, 1] \). As noted in \citet[][p. 80, Eq. (4.2)]{MR2514435}, 
			\( K(x, y) \), 
			being a covariance function, 
			is positive semi-definite and satisfies:
			\begin{align*}
				\int_{0}^{1} \int_{0}^{1} 
				{f(x) \, f(y) \, K(x,y) \, dy \, dx}
				\ge 
				0,
			\end{align*}
			for all $f \in L^{2}$-space.
			This structure ensures that \( L^{2} \) 
			forms a semi-inner product space,
			\citep[see, e.g.,][p. 160]{MR1932358},
			with the semi-inner product:
			\begin{align*}
				\langle
				f, g 
				\rangle
				& = 
				\int_{0}^{1} \int_{0}^{1} 
				{f(x) \, f(y) \, K(x,y) \, dy \, dx}
				\ge 
				0.
			\end{align*}
			Using this semi-inner product,
			the given integrals can be expressed as:
			\begin{align*}
				\Omega_{1} 
				& = 
				\langle 
				f_{1}, f_{1}
				\rangle, 
				\quad 
				\Omega_{2} 
				= 
				\langle 
				f_{1}, f_{1} \, f_{2}
				\rangle, 
				\quad 
				\mbox{and} 
				\quad 
				\Omega_{3} 
				=
				\langle 
				f_{1} \, f_{2}, f_{1} \, f_{2}
				\rangle.
			\end{align*}
			By the Cauchy-Bunyakovsky-Schwarz inequality 
			\citep[see, e.g.,][p. 162]{MR1932358},
			we have:
			\[
			\Omega_{2}^{2}
			\le 
			\Omega_{1} \, \Omega_{2}.
			\]
			Finally, 
			since \( f_{1} \) and \( f_{1} \, f_{2} \) are given 
			to be not linearly dependent, 
			strict inequality holds 
			\citep[see, e.g.,][p. 130]{MR1810041}, 
			giving:
			\[
			\Omega_{2}^{2} < \Omega_{1} \, \Omega_{3},
			\]
			as required.
		\end{proof}
	\end{thm}
	
	\subsection{Kumaraswamy Distribution}
	
	The proposed methodology employs the Kumaraswamy distribution to define observation-specific weights, offering a significant advantages over rigid weighting schemes used in MTM and MWM. 
	To achieve this, 
	we employ weights-generating functions \( J_{j}(u) \) based on Kumaraswamy densities with appropriately chosen parameters. 
	Specifically,
	\( J_{j}(u) \) is defined as:   
	\begin{align}
		\label{eqn:jFun1}
		J_{j}(u) 
		& = 
		\mbox{Kumaraswamy density with appropriate parameters}.
	\end{align}
	For computational simplicity, 
	and in line with the objectives of this study,
	we assume that
	\[
	J_{1}(u) 
	= 
	J_{2}(u) 
	= 
	\cdots 
	=J_{k}(u) 
	\equiv 
	J(u),
	\]
	which implies using identical weights-generating 
	functions for all the $L$-moments. 
	While this assumption simplifies the estimation process,
	it ensures equal weights are applied in the formulations 
	presented in Eqs.~\eqref{eqn:S1} and \eqref{eqn:P1}.
	
\begin{figure}[t!]
		\centering
		\includegraphics[width=0.90\linewidth]
		{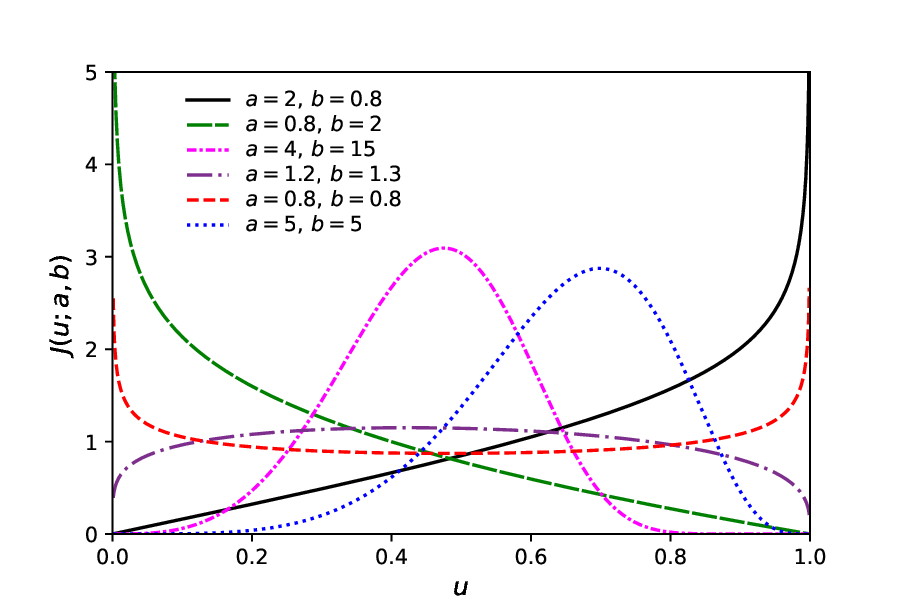}
		\caption{Shapes of the pdf of the
			Kumaraswamy distribution 
			$
			J(u;a,b) 
			= 
			a b u^{a-1} (1 - u^{a})^{b-1},
			$ 
			for $u \in (0, 1)$ 
			and different values of $a$ and $b$.}
		\label{fig:KumDensities}
	\end{figure}

	The probability density function (pdf) of the Kumaraswamy
	random variable is given by 
	\begin{align}
		\label{eqn:KumDen1}
		J(u)
		& \equiv 
		J(u; a, b) 
		= 
		f(u; a, b)
		=
		a \, b \, u^{a-1} \left(1 - u^{a}\right)^{b-1},
		\quad 
		\text{where } 
		u \in (0, 1), \
		a > 0, \
		b > 0,
	\end{align}
	where $a$ and $b$ are the two shape parameters. 
	The parameter $a$ controls the shape of the distribution near 
	$0$ (the lower tail) and parameter $b$ controls the shape
	of the distribution near $1$ (the upper tail). 
	This two-parameter distribution is versatile, accommodating a variety of shapes, including symmetric, skewed, and 
	$U$-shaped distributions. Introduced by \cite{pk80}, the Kumaraswamy distribution provides closed-form expressions for its probability density, cumulative distribution, and quantile functions. Often used as an alternative to the beta distribution, the Kumaraswamy distribution is very popular due to its tractability \cite{MR2655540} and usefulness to develop new generalized distributions such as Kumaraswamy Normal \cite{MR3874077}, Kumaraswamy Laplace \cite{MR3574884}, Kumarswamy Weibull \cite{MR2720934}, among others. 
	In this study we demonstrate that the  Kumaraswamy distribution also has the strength of simplifying computation and enabling a seamless integration into the weighting mechanism. Figure \ref{fig:KumDensities} displays the pdf of  Kumaraswamy distribution for different combinations of shape parameters.
	
\begin{figure}[t!]
		\centering
		\includegraphics[width=0.98\textwidth]{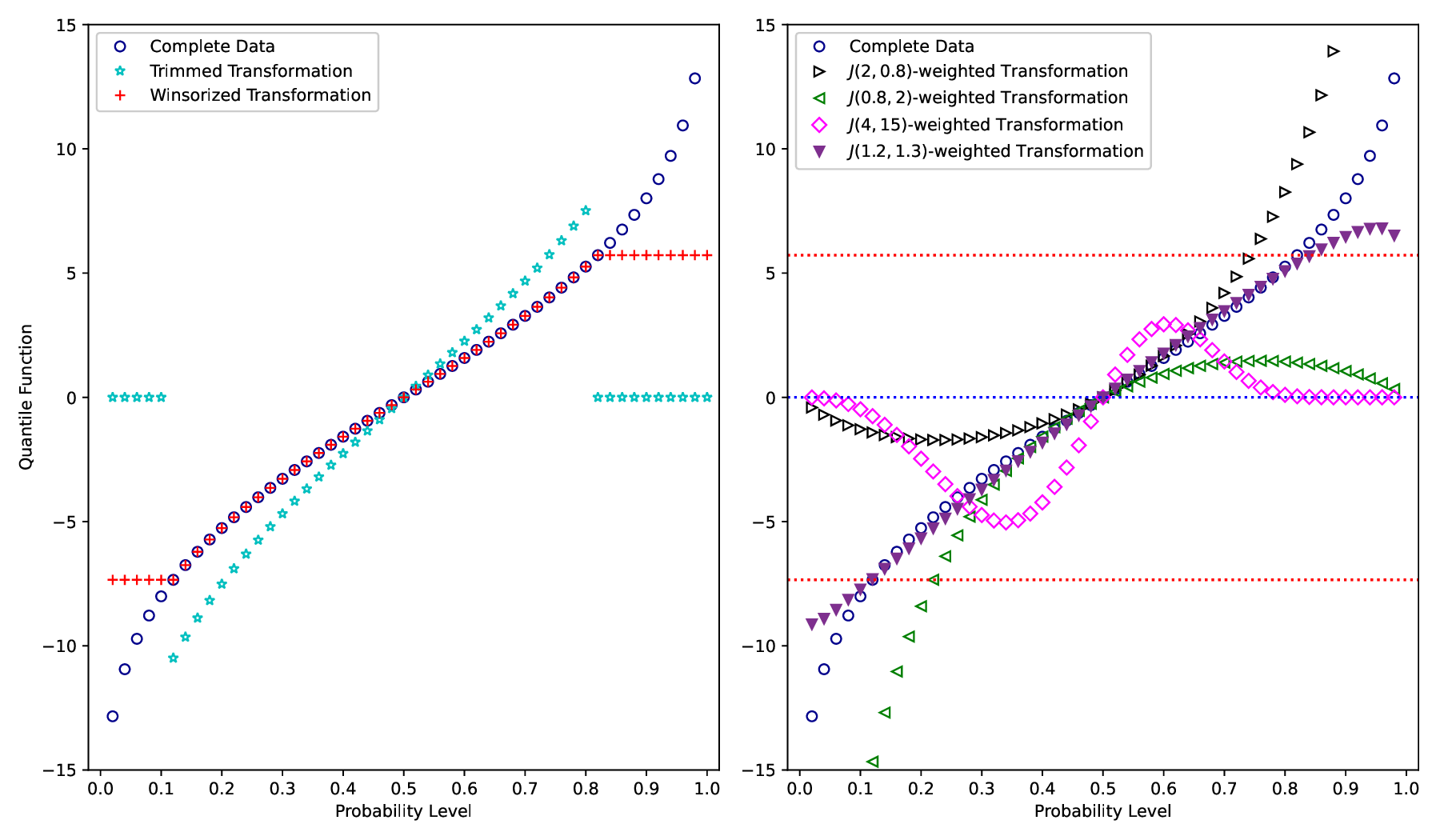}
		\caption{Quantile functions for complete
			data and various transformations. 
			The sample size is $n = 50$. 
			The trimming/winsorizing proportions 
			are 10\% (lower) and 20\% (upper) 
			for the left panel. 
			The parameters for the $J$ 
			function are indicated 
			in the legend on the right panel.}
		\label{fig:TWKTwoColPy1}
	\end{figure}

	To demonstrate the advantages of proposed methods over the MTM and MWM approaches, we have included the quantile functions for complete data and various transformations. The sample size  for this illustration is $n = 50.$
	The trimming and winsorizing proportions are 10\% (lower) and 20\% (upper) for the left panel
	in Figure \ref{fig:TWKTwoColPy1}. The
	parameters for the $J$ function are indicated in the legend on the right panel in Figure
	\ref{fig:TWKTwoColPy1}.
	
	From Figure \ref{fig:KumDensities} and Figure \ref{fig:TWKTwoColPy1} 
	it can be observed that 
	\begin{itemize}
		\item 
		When \(a\) and \(b\) are greater than 1, 
		the density exhibits a bell-shaped behavior. 
		Consequently, if the goal is to assign
		lower weights to the endpoints and 
		higher weights to the central region,
		it is advisable to select \(a > 1\) and \(b > 1\). 
		For example, see 
		\(J(u; a = 4, b = 15)\) -- magenta-colored curve and 
		\(J(u; a = 5, b = 5)\) -- blue-colored curve 
		in Figure \ref{fig:KumDensities}.
		These weighting mechanisms generate smooth weights, 
		as illustrated in Figure \ref{fig:TWKTwoColPy1} 
		(right panel, \(J(4, 15)\) labeled curve).
		
		\item
		When \(a < 1\) and \(b \geq 1\), 
		the density is skewed towards \(0\), 
		assigning heavier weights to the lower-order statistics
		and lighter weights to the higher-order statistics.
		For example,
		see \(J(u; a = 0.8, b = 2)\) -- green-colored curves in
		Figures \ref{fig:KumDensities} 
		and \ref{fig:TWKTwoColPy1} (right panel).
		In contrast, 
		when \(a \geq 1\) and \(b < 1\), 
		the density is skewed towards \(1\), 
		assigning heavier weights to the higher-order statistics 
		and lighter weights to the lower-order statistics.
		For example, 
		see \(J(u; a = 2, b = 0.8)\) -- black-colored curves in 
		Figures \ref{fig:KumDensities} 
		and \ref{fig:TWKTwoColPy1} (right panel).
		This implies that if one intends to assign 
		more weight to lower-order statistics, 
		it is recommended to choose \(a < 1\) and \(b \geq 1\). 
		Conversely, to assign more weight to higher-order statistics,
		it is advisable to choose \(a \geq 1\) and \(b < 1\). 
		
		\item 
		Choosing both \(a > 1\) and \(b > 1\), 
		but close to 1, 
		slightly mitigates the influence of extreme values 
		by assigning lower weights to tail observations 
		and heavier weights to the main body of the sample data, 
		as illustrated by the \(J(1.2, 1.3)\)-labeled curves 
		in Figures \ref{fig:KumDensities} and 
		\ref{fig:TWKTwoColPy1} (right panel).

		\item 
		Finally, by selecting both \(a < 1\) and \(b < 1\), one can assign greater weights to both tails and lighter weights to the middle-order statistics. For example, see \(J(u; a = 0.8, b = 0.8)\) -- red-colored curve in Figure \ref{fig:KumDensities}.
	\end{itemize}
	
	Therefore, by varying the parameters \(a\) and \(b\), 
	one can tailor the density to emphasize specific regions of the data,
	such as assigning heavier weights to central values or shifting focus toward one of the tails. This flexibility is particularly advantageous 
	in claim severity modeling,
	where the underlying distributions often contain outliers,
	enabling the development of stable and robust predictive models. We will us the Kumaraswamy density to facilitate a novel weighting strategy for robust $L$-estimators,
	ensuring to achieve the desired level 
	of efficiency and the robustness.
	Thus, 
	this paper extends the robust $L$-estimation framework
	to incorporate weights generated by Kumaraswamy densities, 
	addressing limitations of MTM and MWM in three key aspects:
	\begin{enumerate}
		\item 
		\textbf{Enhanced Flexibility:} 
		By leveraging the parameterization of the Kumaraswamy distribution, 
		this approach does not completely disregard the trimmed sample 
		observations, 
		as in MTM, nor does it down-weight extreme values uniformly, 
		as in MWM. 
		Instead,
		this method smoothly assigns weights across 
		the entire sample order statistics, 
		enabling the allocation of heavier or lighter weights
		based on the practitioner's preferences or the specific requirements of the scientific problem or business application. 
		
		\item 
		\textbf{Robustness and Efficiency:} 
		The proposed method retains the robustness characteristic of robust $L$-statistics while enhancing efficiency through optimized weighting schemes. 
		This paper offers both theoretical 
		insights and empirical evidence, 
		demonstrating that the proposed robust 
		estimators outperform trimmed and winsorized 
		$L$-estimators as well as MLE, 
		particularly under heavy-tailed and skewed distributions, 
		and when the sample data is contaminated with outliers. 
		
		\item 
		\textbf{Asymptotic Properties:} 
		Building on the foundational asymptotic distributional properties of $L$-statistics established by \cite{MR0203874}, this paper rigorously derives the asymptotic normality of the proposed $L$-estimators. 
		These theoretical advancements provide a robust framework 
		for evaluating the asymptotic relative efficiency (ARE) 
		of the proposed estimators compared to MLE. 
		Through comprehensive theoretical analysis 
		and simulation studies,
		this work highlights the adaptability of
		Kumaraswamy-weighted $L$-estimators to 
		various distributional shapes as 
		illustrated in Figure \ref{fig:TWKTwoColPy1}. 
		The proposed methodology demonstrates resilience against outliers and achieves significant efficiency improvements over MLE, particularly for datasets containing outliers, emphasizing its practical relevance and effectiveness in estimating robust and stable predictive loss models.
	\end{enumerate}
	
	We close this section with the following lemma. 
	
	\medskip 
	
	\begin{lemma}
		\label{lemma:KS_TF1}
		Let $X$ be the Kumaraswamy random variable 
		and $g : (0,1) \to \R$ be a non-degenerate
		continuous function.
		Define $Y := g(X)$. 
		Then being a non-degenerate random variable,
		it immediately follows that 
		$\Var[Y] > 0$.
	\end{lemma}
	
	\section{Parametric Severity Models} 
	\label{sec:ParametricModels}
	
	We now examine the $L$-estimation methodology presented 
	in Section \ref{sec:Methodology} across four parametric examples: 
	the location-scale family, Pareto, lognormal, and Fréchet models.
	The asymptotic performance of the $L$-estimators is evaluated 
	in terms of asymptotic relative efficiency (ARE) 
	compared to the maximum likelihood estimator (MLE). 
	For a scenario involving $k$ parameters, 
	the ARE is defined as follows \citep[see, e.g.,][]{MR595165,MR1652247}:
	\begin{equation} 
		\label{eq:ARE1}
		ARE(\mathcal{C}, MLE) 
		=
		\left( 
		\dfrac{\det
			\left(\bm{\Sigma}_{\text{\tiny MLE}}\right)}
		{\det
			\left(\bm{\Sigma}_{\text{\tiny $\mathcal{C}$}}\right)}
		\right)^{1/k},
	\end{equation}
	where 
	$\bm{\Sigma}_{\text{\tiny MLE}}$
	and 
	$\bm{\Sigma}_{\text{\tiny $\mathcal{C}$}}$
	denote the asymptotic covariance matrices of the MLE and $\mathcal{C}$ estimators, respectively, and $\det$ represents the determinant of a square matrix. 
	The primary rationale for using the MLE as a benchmark lies in its optimal asymptotic performance with respect to variability—granted, of course, that this holds "under certain regularity conditions." For further details, we refer to \citet[][Section 4.1]{MR595165}.
	
	\subsection{Location-scale Families}
	\label{sec:LCF}
	
	Consider
	$X_{1}, X_{2}, \ldots, X_{n} \stackrel{iid}{\sim} X$,
	where $X$ is a location-scale random variable with the CDF
	\begin{align}
		\label{eqn:LC1}
		F(x) 
		& = 
		F_{0} 
		\left( 
		\dfrac{x - \theta}{\sigma}
		\right), 
		\quad 
		-\infty < x < \infty,
	\end{align}
	where $-\infty < \theta < \infty$ 
	and $\sigma > 0$ are, respectively, 
	the location and scale parameters of $X$,
	and $F_{0}$ is the standard parameter-free
	version of $F$, i.e., with $\theta = 0$ and $\sigma = 1$.
	The corresponding percentile/quantile function of $X$ is
	given by
	\begin{align}
		\label{eqn:LC2} 
		F^{-1}(u) 
		& = 
		\theta + \sigma F_{0}^{-1}(u).
	\end{align}
	Since we are estimating two unknown parameters, 
	$\theta$ and $\sigma$, 
	we equate the first two sample $L$-moments with
	their corresponding population $L$-moments.
	Further, knowing 
	$-\infty < \theta < \infty$ and $\sigma > 0$,
	we choose 
	\begin{align}
		\label{eqn:LS_hFun_Def1}
		h_{1}(x)
		& =
		x
		\quad 
		\mbox{and}
		\quad 
		h_{2}(x) 
		= 
		x^{2}.
	\end{align}
	
	From Eq.~\eqref{eqn:P1}, 
	we note that $H_{j} := h_{j} \circ F^{-1}$.
	Then, from Eq.~\eqref{eqn:LC2} and 
	Eq.~\eqref{eqn:LS_hFun_Def1},
	we have 
	\begin{eqnarray*}
		& & 
		H_{1}(u) 
		=
		h_{1}
		\left(
		F^{-1}(u)
		\right) 
		= 
		F^{-1}(u)
		= 
		\theta + \sigma F_{0}^{-1}(u), \\
		\implies 
		& & 
		dH_{1}(u) 
		= 
		\sigma \, d F_{0}^{-1}(u), \\ 
		& & 
		H_{2}(u) 
		=
		h_{2}
		\left(
		F^{-1}(u)
		\right) 
		= 
		\theta^{2} 
		+ 
		2 \theta \sigma \, 
		F_{0}^{-1}(u) 
		+ 
		\sigma^{2} 
		\left[ F_{0}^{-1}(u) \right]^{2}, \\
		\implies & & 
		dH_{2}(u) 
		= 
		2 \theta \sigma \, d F_{0}^{-1}(u)
		+ 
		2 \sigma^{2} F_{0}^{-1}(u) \, 
		dF_{0}^{-1}(u).
	\end{eqnarray*}
	
	From Eq.~\eqref{eqn:S1}, 
	the first two sample $L$-moments 
	are given by 
	\begin{align}
		\label{eqn:S2}
		\begin{cases}
			\displaystyle 
			\widehat{\mu}_{1} 
			=
			\dfrac{1}{n}
			\sum_{i=1}^{n}
			J \left( \dfrac{i}{n+1} \right) 
			h_{1}(X_{i:n}) 
			= 
			\dfrac{1}{n}
			\sum_{i=1}^{n}
			J \left( \dfrac{i}{n+1} \right) 
			X_{i:n}, \\[15pt]
			\displaystyle 
			\widehat{\mu}_{2} 
			=
			\dfrac{1}{n}
			\sum_{i=1}^{n}
			J \left( \dfrac{i}{n+1} \right) 
			h_{2}(X_{i:n}) 
			=
			\dfrac{1}{n}
			\sum_{i=1}^{n}
			J \left( \dfrac{i}{n+1} \right) 
			X_{i:n}^{2}.
		\end{cases}
	\end{align}
	
	The corresponding first two population 
	$L$-moments using Eq.~\eqref{eqn:P1}
	takes the form
	\begin{align}
		\label{eqn:P2}
		\begin{cases}
			\mu_{1}
			\equiv 
			\mu_{1} 
			\left( 
			\theta, \sigma
			\right)  
			= 
			\int_{0}^{1}
			J(u)H_{1}(u) \, du
			= 
			\int_{0}^{1}
			J(u)F^{-1}(u) \, du
			= 
			\theta + \sigma c_{1}, \\[10pt]
			\mu_{2}
			\equiv 
			\mu_{2} 
			\left( 
			\theta, \sigma
			\right)  
			= 
			\int_{0}^{1}
			J(u)H_{2}(u) \, du
			= 
			\int_{0}^{1}
			J(u)
			\left[ F^{-1} (u) \right]^{2} \, du
			= 
			\theta^{2} 
			+ 
			2 \theta \sigma c_{1} 
			+ 
			\sigma^{2} c_{2},
		\end{cases}
	\end{align}
	where 
	\begin{align}
		\label{eqn:ConstantC1}
		c_{k} 
		& \equiv 
		c_{k}\left( F_{0}, J \right)
		=
		\int_{0}^{1} J(u) 
		\left[ F_{0}^{-1} (u) \right]^{k} du, 
		\quad 
		k = 1, 2.
	\end{align}
	
	Since $c_{1}$ and $c_{2}$ do not depend on
	the parameters to be estimated, 
	equating 
	$\mu_{1} 
	= 
	\widehat{\mu}_{1}$ and $\mu_{2} 
	= 
	\widehat{\mu}_{2}
	$ 
	yields an explicit system of equations 
	for $\theta$ and $\sigma$ as 
	\begin{align}
		\label{eqn:TS1}
		\left\{
		\begin{array}{lll}
			\widehat{\theta}_{\mbox{\tiny K}}
			& = &
			\widehat{\mu}_{1} - c_{1} \widehat{\sigma}_{\mbox{\tiny K}} 
			=: 
			g_{1}(\widehat{\mu}_{1},\widehat{\mu}_{2}), \\[10pt]
			\widehat{\sigma}_{\mbox{\tiny K}}
			& = & 
			\sqrt{(\widehat{\mu}_{2}-\widehat{\mu}_{1}^{2})/\eta} =: 
			g_{2}(\widehat{\mu}_{1},\widehat{\mu}_{2}),
			\quad 
			\text{where}
			\quad 
			\eta 
			\equiv 
			\eta
			\left( 
			F_{0}, J
			\right) 
			: = 
			c_{2}-c_{1}^{2}.
		\end{array}
		\right.
	\end{align}
	
	\medskip 
	
	\begin{note}
		\label{note:M2MinusMu1Neg1}
		In estimating $\widehat{\sigma}_{\text{\tiny K}}$, 
		it is possible that 
		$\widehat{\mu}_{2} - \widehat{\mu}_{1}^{2} < 0$ 
		for certain combinations of an observed 
		sample (small sample size) and the 
		weight-generating function $J$.
		In such cases,
		we may need to change the weights-generating 
		function $J$ which will produce  
		$\widehat{\mu}_{2} - \widehat{\mu}_{1}^{2} > 0$.
		
		To illustrate this fact, 
		we consider consider a toy 
		sample dataset:
		$
		X  = (1, 2, 3, 4, 5). 
		$
		Also consider $a = 5 = b$.
		Then it follows that 
		\begin{align*}
			\widehat{\mu}_{1} 
			& = 
			\sum_{i=1}^{5} 
			J
			\left( 
			\dfrac{i}{6}
			\right) 
			X_{i:5}
			=
			4.7398
			\quad 
			\mbox{and}
			\quad 
			\widehat{\mu}_{2} 
			= 
			\sum_{i=1}^{5} 
			J
			\left( 
			\dfrac{i}{6}
			\right) 
			X_{i:5}^{2}
			=
			19.4222,
		\end{align*}
		giving us 
		\[
		\widehat{\mu}_{2} 
		- 
		\widehat{\mu}_{1}^{2}
		=
		-3.0437
		< 
		0. 
		\]
		
		In this scenario, 
		the estimation formula for 
		$\widehat{\sigma}_{\mbox{\tiny K}}$
		as given in Eq.~\eqref{eqn:TS1} 
		is deemed unsuitable.
		\qed 
	\end{note}
	
	From Theorem \ref{thm:CGJ1}, 
	the entries of the variance-covariance matrix
	$\Sigma$
	evaluated using Eq.~\eqref{eqn:mtm_var_cov3} are 
	\begin{eqnarray*}
		\sigma_{11}^{2}
		& = &
		\int_{0}^{1} 
		\int_{0}^{1} 
		J(v) J(w)
		K(v,w) \, 
		dH_{1}(v) \, 
		dH_{1}(w) \\
		& = &
		\sigma^{2} 
		\int_{0}^{1} 
		\int_{0}^{1} 
		J(v) J(w)
		K(v,w) \, 
		dF_{0}^{-1}(v) \, 
		dF_{0}^{-1}(w) \\ 
		& = & 
		\sigma^{2} \Lambda_{1}, \\
		\sigma_{12}^{2}
		& = &
		\int_{0}^{1} 
		\int_{0}^{1} 
		J(v) J(w)
		K(v,w) \, 
		dH_{1}(v) \, 
		dH_{2}(w) \\
		& = &
		2 \theta \sigma^{2} 
		\int_{0}^{1} 
		\int_{0}^{1} 
		J(v) J(w)
		K(v,w) \, 
		dF_{0}^{-1}(v) \, 
		dF_{0}^{-1}(w) \\
		& & 
		+ 
		2 \sigma^{3} 
		\int_{0}^{1} 
		\int_{0}^{1} 
		J(v) J(w)
		K(v,w) \, 
		F_{0}^{-1}(w) \, 
		dF_{0}^{-1}(v) \,
		dF_{0}^{-1}(w) \\ 
		& = & 
		2 \theta \sigma^{2} \Lambda_{1}
		+ 
		2 \sigma^{3} \Lambda_{2}, \\
		\sigma_{22}^{2}
		& = &
		\int_{0}^{1} 
		\int_{0}^{1} 
		J(v) J(w)
		K(v,w) \, 
		dH_{2}(v) \, 
		dH_{2}(w) \\
		& = &
		4 \theta^{2} \sigma^{2}
		\int_{0}^{1} 
		\int_{0}^{1} 
		J(v) J(w)
		K(v,w) \,  
		dF_{0}^{-1}(v) \, 
		dF_{0}^{-1}(w) \\
		& & 
		+ 
		8 \theta \sigma^{3} 
		\int_{0}^{1} 
		\int_{0}^{1} 
		J(v) J(w)
		K(v,w) \, 
		F_{0}^{-1}(w) \, 
		dF_{0}^{-1}(v) \, 
		dF_{0}^{-1}(w) \\
		& & 
		+ 
		4 \sigma^{4} 
		\int_{0}^{1} 
		\int_{0}^{1} 
		J(v) J(w)
		K(v,w) \, 
		F_{0}^{-1}(v) \, 
		F_{0}^{-1}(w) \, 
		dF_{0}^{-1}(v) \, 
		dF_{0}^{-1}(w) \\
		& = & 
		4 \theta^{2} \sigma^{2} \Lambda_{1} 
		+ 
		8 \theta \sigma^{3} \Lambda_{2} 
		+ 
		4 \sigma^{4} \Lambda_{3},
	\end{eqnarray*}
	where the integral notations 
	\begin{eqnarray*}
		\Lambda_{1}
		& \equiv & 
		\Lambda_{1}
		\left(
		F_{0}, J
		\right)
		= 
		\int_{0}^{1} 
		\int_{0}^{1} 
		J(v) J(w)
		K(v,w) \, 
		dF_{0}^{-1}(v) \, 
		dF_{0}^{-1}(w), \\
		\Lambda_{2}
		& \equiv & 
		\Lambda_{2}
		\left(
		F_{0}, J
		\right)
		=  
		\int_{0}^{1} 
		\int_{0}^{1} 
		J(v) J(w)
		K(v,w) \, 
		F_{0}^{-1}(w) \, 
		dF_{0}^{-1}(v) \,
		dF_{0}^{-1}(w), \\
		\Lambda_{3}
		& \equiv & 
		\Lambda_{3}
		\left(
		F_{0}, J
		\right)
		= 
		\int_{0}^{1} 
		\int_{0}^{1} 
		J(v) J(w)
		K(v,w) \, 
		F_{0}^{-1}(v) \, 
		F_{0}^{-1}(w) \, 
		dF_{0}^{-1}(v) \, 
		dF_{0}^{-1}(w),
	\end{eqnarray*}
	do not depend on the parameters to be estimated.
	
	\begin{cor}
		\label{cor:LSCov1}
		The following inequalities hold:
		\begin{itemize}
			\item[(i)] 
			$\eta = c_{2} - c_{1}^{2} > 0$.
			
			\item[(ii)] 
			\( 
			\Lambda_{1} \, \Lambda_{3} 
			-
			\Lambda_{2}^{2}
			> 
			0.
			\)
		\end{itemize}
		
		\begin{proof}
			We use Theorem \ref{thm:GeneralCSIneq1}
			and Lemma \ref{lemma:KS_TF1} to establish 
			the required inequalities. 
			\begin{itemize} 
				\item[(i)]
				This immediately follows by taking 
				\( 
				g(x) = F_{0}^{-1}(x)
				\) 
				in Lemma \ref{lemma:KS_TF1}.
				
				\item[(ii)]
				This follows from Theorem \ref{thm:GeneralCSIneq1}
				with the following assignments:
				\begin{align*}
					f_{1}(x) 
					& =  
					\dfrac{J(x)}
					{
						f_{0} 
						\left( 
						F_{0}^{-1}(x)
						\right) 
					}
					\quad 
					\mbox{and}
					\quad 
					f_{2}(x) 
					= 
					F_{0}^{-1}(x).
					\qedhere 
				\end{align*}
			\end{itemize}
		\end{proof}
	\end{cor}
	
	Similarly, from Theorem \ref{thm:CGJ2},
	the entries of the matrix $\bm {D}$
	are obtained by differentiating 
	the functions $g_{j}$ defined in Eq. \eqref{eqn:TS1}:
	\begin{eqnarray}
		\label{eqn:LSmD1}
		{\bm D} 
		& = & 
		\left[d_{ij}\right]_{2\times 2}
		=
		\left[\left. \frac{\partial g_{i}}{\partial \widehat{\mu}_{j}}\right\vert_{\widehat{\bm{\mu}}
			=
			\bm{\mu}}\right]_{2 \times 2} 
		= 
		\dfrac{1}{\sigma \eta}
		\begin{bmatrix}
			c_{1} \theta + c_{2} \sigma & 
			-0.5 c_{1} \\[5pt]
			- \theta - c_{1} \sigma & 
			0.5
		\end{bmatrix}.
	\end{eqnarray}
	
	Therefore, it follows that 
	\begin{align}
		\label{eqn:ThetaAsym1}
		\left( 
		\widehat{\theta}_{\mbox{\tiny K}}, 
		\widehat{\sigma}_{\mbox{\tiny K}}
		\right) 
		& \sim  
		\mathcal{AN}
		\left( 
		\left(\theta, \sigma \right), 
		\dfrac{1}{n} \bm{S}_{\mbox{\tiny K}}
		\right),
	\end{align}
	where 
	\begin{align}
		\bm{S}_{\mbox{\tiny K}}
		& = 
		\bm{D} \bm{\Sigma} \bm{D}' 
		\nonumber \\
		& =
		\frac{\sigma^{2} }
		{\eta^2}
		\begin{bmatrix}
			\Lambda_1 c_{2}^{2} 
			- 2c_1 c_2 \Lambda_2 
			+ c_{1}^{2} \Lambda_3 & 
			-\Lambda_1 c_{1} c_{2} 
			+ c_2 \Lambda_2 
			+ c_{1}^{2} \Lambda_2 - c_1 \Lambda_3 \\[5pt]
			-\Lambda_1 c_{1} c_{2} 
			+ c_2 \Lambda_2
			+ c_{1}^{2} \Lambda_2
			- c_1 \Lambda_3 & 
			\Lambda_1 c_{1}^{2} 
			- 2c_1 \Lambda_2
			+ \Lambda_3
		\end{bmatrix}. 
		\label{eqn:SMat1}
	\end{align}
	
	From Eq.~\eqref{eqn:SMat1}, 
	it follows that 
	\begin{align}
		\label{eqn:SMat2}
		\mbox{det}
		\left( 
		\bm{S}_{\mbox{\tiny K}}
		\right) 
		& = 
		\dfrac{
			\sigma^{4} 
			\left(
			\Lambda_{1} \, \Lambda_{3}
			-
			\Lambda_{2}^{2}
			\right)
		}
		{\eta^{2}}.
	\end{align}
	
	From Corollary \ref{cor:LSCov1}, 
	it follows that 
	\( 
	\mbox{det}
	\left( 
	\bm{S}_{\mbox{\tiny K}}
	\right)
	>
	0.
	\)
	
	\subsection{Pareto Severity Model}
	
	Here we develop the methodology for a 
	single parameter Pareto model. 
	The CDF of the single parameter Pareto 
	random variable $X \sim \mbox{Pareto I}(\alpha, x_{0})$
	is given by 
	\begin{align}
		\label{eqn:P1CDF1}
		F(x) 
		& = 
		1 
		- 
		\left( 
		\dfrac{x}{x_{0}}
		\right)^{-\alpha}, 
		\quad 
		x > x_{0},
	\end{align}
	where $\alpha > 0$ is the shape parameter
	and $x_{0} > 0$ is assumed to be know. 
	Clearly, from Eq. \eqref{eqn:P1CDF1}, 
	the corresponding quantile function is
	\begin{align}
		\label{eqn:P1QF1}
		F^{-1}(u) 
		& = 
		x_{0}(1-u)^{-1/\alpha}.
	\end{align}
	
	Since the distribution $F$ contains only one unknown parameter, 
	it is sufficient to use a single $L$-moment for estimation, 
	and we choose
	$h(x) = \log{(x/x_{0})}$. 
	
	With the chosen $h$-function, 
	Eqs. \eqref{eqn:S1} and \eqref{eqn:P1}
	respectively take the following form
	\begin{eqnarray}
		\widehat{\mu} 
		& = & 
		\dfrac{1}{n}
		\sum_{i=1}^{n}
		J \left( \dfrac{i}{n+1} \right) 
		\log{(X_{i:n}/x_{0})}, 
		\label{eqn:P1S1} \\
		\mu
		& = & 
		\int_{0}^{1}
		J(u)H(u) \, du 
		= 
		ab 
		\int_{0}^{1} 
		u^{a-1} \left( 1- u^{a} \right)^{b-1}
		\log{\left( F^{-1}(u)/x_{0} \right)} \, du 
		\nonumber \\ 
		& = & 
		- 
		\dfrac{ab}{\alpha} 
		\int_{0}^{1} 
		u^{a-1} \left( 1- u^{a} \right)^{b-1}
		\log{\left( 1 - u \right)} \, du 
		\nonumber \\ 
		& = & 
		- 
		\dfrac{I_{1}(a,b)}{\alpha},
		\quad 
		I_{1}
		\equiv 
		I_{1}(a,b) 
		:= 
		\int_{0}^{1} 
		J(u)
		\log{\left( 1 - u \right)} \, du. 
		\label{eqn:P1P1}
	\end{eqnarray}
	
	\begin{note} 
		By using the binomial series expansion for 
		$\left( 1 - u^{a} \right)^{b-1}$, 
		we can express the population mean $\mu$
		from Eq. \eqref{eqn:P1P1} in terms
		of hypergeometric function as follows: 
		\begin{align}
			\mu 
			& = 
			\dfrac{b}{\alpha}
			\sum_{n=0}^{\infty} 
			\dfrac{(-1)^{n} \binom{b-1}{n}}
			{n + 1} \
			{}_{2}F_{1} 
			\left( 
			1, a (n+1); a (n+1) + 1; 1
			\right).
			\label{eqn:HG1}
		\end{align}
		But for computational purpose,
		Eq. \eqref{eqn:P1P1} would be easier 
		by using numerical techniques 
		rather than the form given by Eq. \eqref{eqn:HG1}.
		\qed 
	\end{note}
	
	Now, setting 
	$\mu = \widehat{\mu}$ 
	gives us 
	\begin{align}
		\widehat{\alpha}_{\mbox{\tiny K}}
		& = 
		- 
		\dfrac{I_{1}(a, b)}{\widehat{\mu}} 
		=: 
		g_{1}
		\left( 
		\widehat{\mu}
		\right). 
	\end{align}
	
	For the asymptotic behavior of 
	$\widehat{\alpha}_{\mbox{\tiny K}}$,
	the entries of the matrices 
	$\bm{\Sigma}$ and $\bm{D}$,
	from Theorem \ref{thm:CGJ1} 
	and Theorem \ref{thm:CGJ2}, 
	respectively,
	for a single dimension are now calculated as:
	\begin{align*}
		\sigma_{11}^{2} 
		& = 
		\int_{0}^{1} 
		\int_{0}^{1} 
		J(v) J(w)
		K(v,w) \, 
		dH(v) \, 
		dH(w) \\
		& = 
		\dfrac{1}{\alpha^{2}} 
		\int_{0}^{1} 
		\int_{0}^{1} 
		\dfrac{J(v) J(w) K(v,w)}
		{\ol{v} \, \ol{w}} \, dv \, dw \\
		& = 
		\dfrac{I_{2}(a,b)}{\alpha^{2}} \\ 
		{\bm D} 
		& = 
		\left[d_{ij}\right]_{1 \times 1} 
		=
		\left[\left. \frac{\partial g_{1}}{\partial \widehat{\mu}_{1}}\right\vert_{\widehat{\bm{\mu}}
			=
			\bm{\mu}}\right]_{1 \times 1} 
		=
		\dfrac{\alpha^{2}}{I_{1}(a, b)}.
	\end{align*}
	
	Therefore, it follows that 
	\begin{align}
		\label{eqn:AlphaAsym1}
		\widehat{\alpha}_{\mbox{\tiny K}}
		& \sim  
		\mathcal{AN}
		\left( 
		\alpha, 
		\dfrac{\alpha^{2}}{n} S_{\mbox{\tiny K}}
		\right),
		\quad 
		S_{\mbox{\tiny K}}
		=
		\dfrac{I_{2}(a,b)}
		{I_{1}^{2}(a,b)}.
	\end{align}
	
	We note that 
	$
	\widehat{\alpha}_{\mbox{\tiny MLE}}
	\sim 
	\mathcal{AN}
	\left( 
	\alpha, 
	\dfrac{\alpha^{2}}{n}
	\right).
	$
	Thus, from Eq.~\eqref{eq:ARE1} and 
	Eq.~\eqref{eqn:AlphaAsym1}, 
	we have 
	\begin{align}
		\label{eqn:PI_K_MLE_ARE}
		\mbox{ARE}
		\left( 
		\widehat{\alpha}_{\mbox{\tiny K}},
		\widehat{\alpha}_{\mbox{\tiny MLE}}
		\right)
		& =
		\dfrac{1}{S_{\mbox{\tiny K}}}
		=
		\dfrac{I_{1}^{2}(a,b)}{I_{2}(a,b)}.
	\end{align}
	
	\begin{table}[hbt!]
		\caption{  
			$\mbox{ARE}\left( 
			\widehat{\alpha}_{\mbox{\tiny K}},
			\widehat{\alpha}_{\mbox{\tiny MLE}}
			\right)$
			for selected values of $a$ and $b$.
		}
		\label{table:ParetoARE}
		\centering
		\begin{tabular}{|c|cccccccccc|}
			\hline 
			\multirow{2}{*}{$a$} & 
			\multicolumn{10}{|c|}{$b$} \\ 
			\cline{2-11} 
			{} & 0.3 & 0.5 & 0.8 & 1 & 1.3 & 2 & 5 & 7 & 15 & 20 \\
			\hline 
			0.3  & 0.143 & 0.509 & 0.973 & 0.940 & 0.775 & 0.459 & 0.088 & 0.041 & 0.006 & 0.003 \\
			0.5  & 0.140 & 0.490 & 0.972 & 0.975 & 0.851 & 0.572 & 0.170 & 0.100 & 0.028 & 0.016 \\
			0.8  & 0.135 & 0.464 & 0.956 & 0.997 & 0.920 & 0.693 & 0.289 & 0.201 & 0.084 & 0.060 \\
			1.0  & 0.133 & 0.449 & 0.941 & 1.000 & 0.947 & 0.750 & 0.360 & 0.265 & 0.129 & 0.097 \\
			1.2  & 0.130 & 0.436 & 0.925 & 0.998 & 0.964 & 0.794 & 0.422 & 0.325 & 0.175 & 0.138 \\
			2.0  & 0.122 & 0.392 & 0.859 & 0.964 & 0.982 & 0.891 & 0.601 & 0.509 & 0.344 & 0.297 \\
			4.0  & 0.108 & 0.325 & 0.728 & 0.855 & 0.924 & 0.928 & 0.787 & 0.725 & 0.596 & 0.552 \\
			5.0  & 0.103 & 0.302 & 0.680 & 0.808 & 0.886 & 0.915 & 0.820 & 0.771 & 0.662 & 0.623 \\
			7.0  & 0.096 & 0.268 & 0.604 & 0.729 & 0.818 & 0.874 & 0.842 & 0.812 & 0.736 & 0.707 \\
			10.0 & 0.087 & 0.234 & 0.524 & 0.642 & 0.734 & 0.809 & 0.831 & 0.819 & 0.778 & 0.760 \\
			\hline 
		\end{tabular}
	\end{table}
	
\begin{figure}[b!]
		\centering
		\includegraphics[width=0.75\textwidth,height=0.50\textheight]{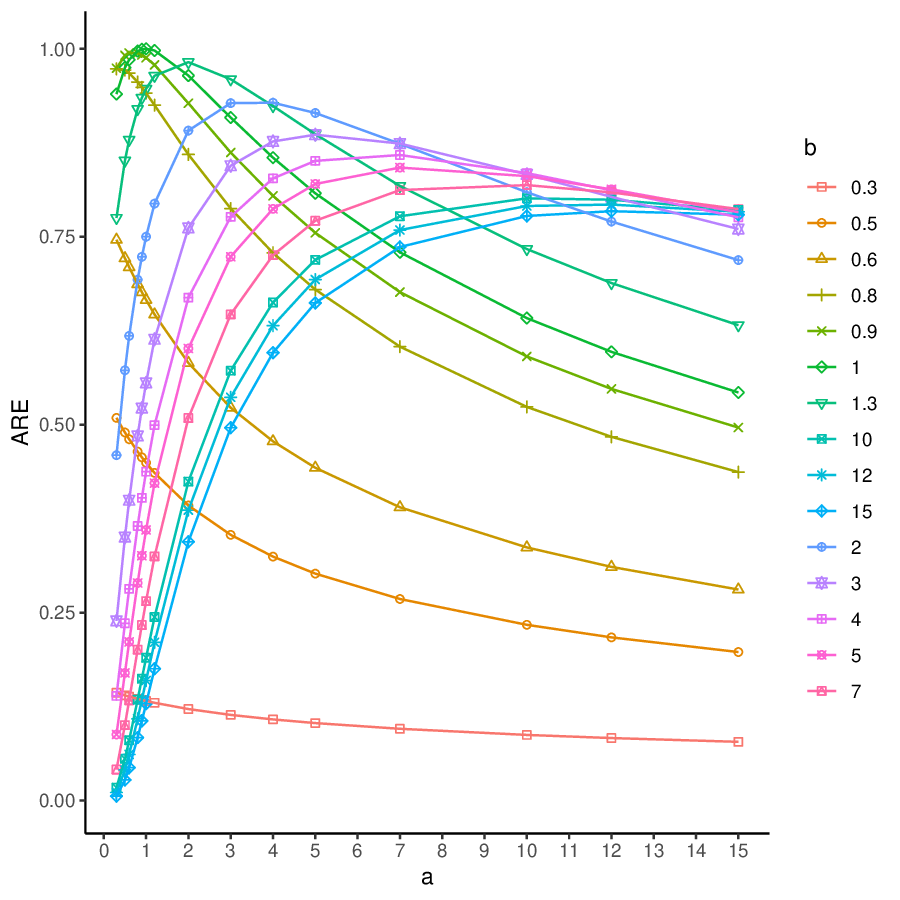} 
		\caption{ARE values of the Pareto severity 
			model presented as interactions between 
			the shape parameters \( a \) and \( b \).}
		\label{fig:ParetoARE}
	\end{figure}

	We now examine the efficiency of our proposed method of using Kumaraswamy density weights with respect to their MLE estimators using Eq. \eqref{eq:ARE1}. 
	The numerical values of these AREs, calculated using Eq.~\eqref{eqn:PI_K_MLE_ARE}, are provided in Table~\ref{table:ParetoARE}, with the corresponding interaction plot presented in Figure~\ref{fig:ParetoARE} for various combinations of the parameters \( a \) and \( b \). From both Table~\ref{table:ParetoARE} and Figure~\ref{fig:ParetoARE}, it is evident that maximum efficiency is achieved when both \( a \) and \( b \) are close to 1. This weighting mechanism yields lighter tails compared to the observed sample data, effectively controlling the influence of smaller or larger order statistics.
	
	Notably, when \( a = 1 = b \), the \( J \)-weighted \( L \)-estimator simplifies to the method of moments (MM). For a single-parameter Pareto distribution, the MM and MLE are identical, resulting in an efficiency of 1.
	
	Additionally, the AREs exhibit a threshold-like behavior around 
	\( (a = 1, b = 1) \). Specifically, if a practitioner is willing to tolerate an ARE level of \( 75\% \) or higher, there are numerous combinations of \( a \) and \( b \) that provide substantial flexibility in selecting an appropriate weighting mechanism. These combinations facilitate the development of more stable fitted models that remain robust against perturbations in the underlying models. At the same time, they maintain the desired efficiency level and ensure that weights are assigned smoothly—unlike the constant weighting mechanisms employed by MTM and MWM—across the relevant portions of the observed sample data.
	
	\subsection{Lognormal Severity Model}
	
	Consider $X_{1}, X_{2}, \ldots, X_{n} \stackrel{iid}{\sim} X$,
	where $X$ has a lognormal distribution function
	\begin{align}
		\label{eqn:LN1}
		F(x) 
		& = 
		\Phi 
		\left( 
		\dfrac{\log{(x-x_{0})} - \theta}{\sigma}
		\right) 
		\quad 
		x > x_{0},
	\end{align}
	where $-\infty < \theta < \infty$ 
	and $\sigma > 0$ and $\Phi$ is the 
	standard normal cdf. 
	Since $X$ is lognormal, 
	then it is well know that 
	$\log{(X-x_{0})}$ is normal,
	a member of the location-scale family. 
	So, the results of Section \ref{sec:LCF}
	are applicable with the two functions:
	\[
	h_{1}(x)
	= 
	\log{(x - x_{0})}
	\quad 
	\mbox{and} 
	\quad 
	h_{2}(x) 
	= 
	\left( 
	\log{(x - x_{0})}
	\right)^{2}.
	\]
	
	That is, the first two sample 
	$L$-moments corresponding to 
	Eq. \eqref{eqn:S2} are given by 
	\begin{align}
		\label{eqn:S3}
		\begin{cases}
			\displaystyle 
			\widehat{\mu}_{1} 
			= 
			\dfrac{1}{n}
			\sum_{i=1}^{n}
			J \left( \dfrac{i}{n+1} \right) 
			\log{ 
				\left(
				X_{i:n} - x_{0} 
				\right)
			}, \\[15pt]
			\displaystyle 
			\widehat{\mu}_{2} 
			=
			\dfrac{1}{n}
			\sum_{i=1}^{n}
			J \left( \dfrac{i}{n+1} \right) 
			\left( 
			\log{ 
				\left(
				X_{i:n} - x_{0} 
				\right)
			}
			\right)^{2}.
		\end{cases}
	\end{align}
	
	The corresponding first two population 
	$L$-moments are given by 
	Eq.~\eqref{eqn:P2} with the 
	developed $H_{1}$ and $H_{2}$ functions. 
	Finally, the estimators 
	$\widehat{\theta}_{\text{\tiny K}}$
	and 
	$\widehat{\sigma}_{\text{\tiny K}}$
	are solved as in Eq.~\eqref{eqn:TS1}, 
	where the expressions for 
	$\widehat{\mu}_{1}$
	and 
	$\widehat{\mu}_{2}$
	are found in Eq.~\eqref{eqn:S3}.
	From Eq.~\eqref{eqn:ThetaAsym1}, 
	the asymptotic distribution is given by 
	\begin{align}
		\label{eqn:ThetaAsym2}
		\left( 
		\widehat{\theta}_{\mbox{\tiny K}}, 
		\widehat{\sigma}_{\mbox{\tiny K}}
		\right) 
		& \sim  
		\mathcal{AN}
		\left( 
		\left(\theta, \sigma \right), 
		\dfrac{1}{n} \bm{S}_{\mbox{\tiny K}}
		\right),
	\end{align}
	with $\bm{S}_{\mbox{\tiny K}}$ given by 
	Eq.~\eqref{eqn:SMat1},
	however, in this case, 
	we use the standard normal distribution 
	\(F_{0}(u) = \Phi(u)\) 
	rather than the standardized location-scale 
	distribution employed previously.
	
	From \cite{MR1987777},
	we note that 
	the maximum likelihood estimated
	parameters are given by
	\begin{align}
		\label{eqn:MLELN1}
		\left\{
		\begin{array}{lll}
			\widehat{\theta}_{\mbox{\tiny MLE}}
			& = &
			\dfrac{1}{n} 
			\displaystyle 
			\sum_{i=1}^{n} \log{\left( X_{i} - x_{0} \right)}, \\[15pt]
			\widehat{\sigma}_{\mbox{\tiny MLE}}
			& = & 
			\sqrt{
				\dfrac{1}{n} 
				\displaystyle 
				\sum_{i=1}^{n} 
				\left( 
				\log{(X_{i} - x_{0})}
				-
				\widehat{\theta}_{\mbox{\tiny MLE}}
				\right)^{2}
			}.
		\end{array}
		\right.
	\end{align}
	
	\begin{table}[t!]
		\caption{ 
			$
			\mbox{ARE}
			\left( 
			\left(\widehat{\theta}_{\mbox{\tiny K}},
			\widehat{\sigma}_{\mbox{\tiny K}}\right),
			\left(\widehat{\theta}_{\mbox{\tiny MLE}},\widehat{\sigma}_{\mbox{\tiny MLE}}\right)
			\right)$
			for selected values of $a$ and $b$.
		}
		\label{table:LNARE}
		\centering
		\begin{tabular}{|c|cccccccccc|}
			\hline 
			\multirow{2}{*}{$a$} & 
			\multicolumn{10}{|c|}{$b$} \\ 
			\cline{2-11} 
			{} & 0.3 & 0.5 & 0.8 & 1 & 1.3 & 2 & 5 & 7 & 15 & 20 \\
			\hline 
			0.3  & 0.142 & 0.264 & 0.204 & 0.152 & 0.103 & 0.053 & 0.017 & 0.014 & 0.013 & 0.014 \\
			0.5  & 0.214 & 0.538 & 0.582 & 0.479 & 0.356 & 0.203 & 0.054 & 0.033 & 0.012 & 0.009 \\
			0.8  & 0.176 & 0.549 & 0.962 & 0.950 & 0.846 & 0.623 & 0.262 & 0.184 & 0.080 & 0.058 \\
			1.0  & 0.152 & 0.479 & 0.950 & 1.000 & 0.955 & 0.782 & 0.412 & 0.314 & 0.164 & 0.128 \\
			1.2  & 0.133 & 0.420 & 0.892 & 0.977 & 0.974 & 0.853 & 0.518 & 0.417 & 0.247 & 0.201 \\
			2.0  & 0.092 & 0.279 & 0.662 & 0.782 & 0.847 & 0.844 & 0.672 & 0.599 & 0.449 & 0.401 \\
			4.0  & 0.057 & 0.155 & 0.389 & 0.489 & 0.565 & 0.622 & 0.608 & 0.584 & 0.520 & 0.495 \\
			5.0  & 0.050 & 0.128 & 0.323 & 0.412 & 0.483 & 0.544 & 0.555 & 0.541 & 0.499 & 0.482 \\
			7.0  & 0.040 & 0.096 & 0.242 & 0.314 & 0.376 & 0.435 & 0.467 & 0.464 & 0.446 & 0.437 \\
			10.0 & 0.033 & 0.071 & 0.177 & 0.233 & 0.283 & 0.336 & 0.377 & 0.380 & 0.377 & 0.374 \\
			\hline 
		\end{tabular}
	\end{table}
	
	\begin{figure}[hbt!]
		\centering
		\includegraphics[width=0.75\textwidth,height=0.50\textheight]{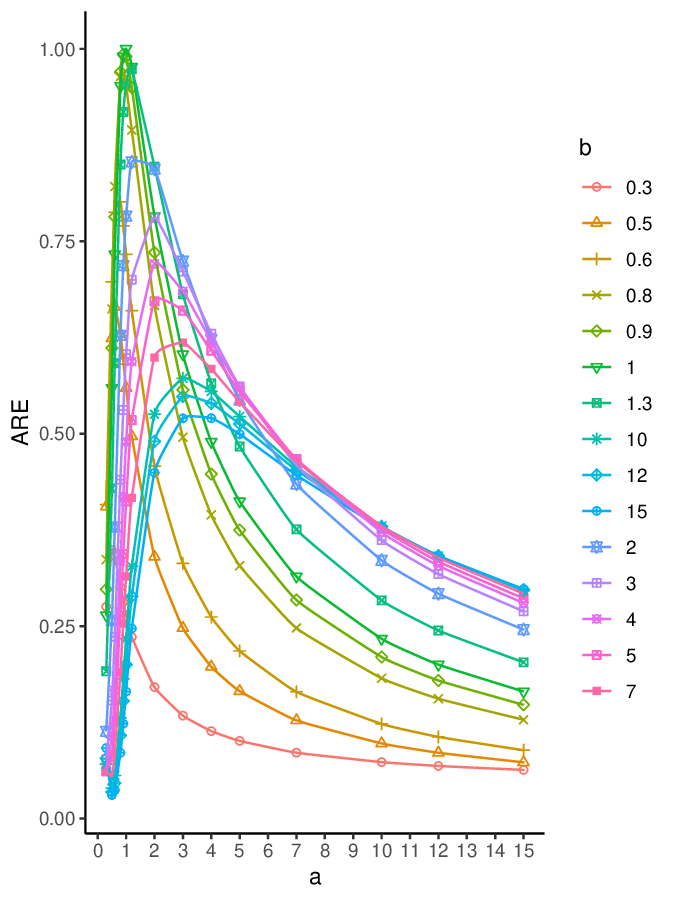} 
		\caption{ARE values of the lognormal severity 
			model presented as interactions between 
			the shape parameters \( a \) and \( b \).}
		\label{fig:LN_ARE}
	\end{figure}
	
	Further, it follows that 
	\begin{align}
		\label{eqn:LNThetaAsym1}
		\left( 
		\widehat{\theta}_{\mbox{\tiny MLE}}, 
		\widehat{\sigma}_{\mbox{\tiny MLE}}
		\right) 
		& \sim  
		\mathcal{AN}
		\left( 
		\left(\theta, \sigma \right), 
		\dfrac{1}{n} \bm{S}_{\mbox{\tiny MLE}}
		\right), 
		\mbox{ where } 
		\bm{S}_{\mbox{\tiny MLE}}
		= 
		\sigma^{2}
		\begin{bmatrix}
			1 & 0 \\ 
			0 & 0.5
		\end{bmatrix}
		\mbox{ with }
		\mbox{det}
		\left(
		\bm{S}_{\mbox{\tiny MLE}}
		\right) 
		=
		\dfrac{\sigma^{4}}{2}.
	\end{align}
	
	Thus, from
	Eqs.~\eqref{eq:ARE1},
	\eqref{eqn:SMat2}, and 
	\eqref{eqn:LNThetaAsym1}, 
	we have
	\begin{align}
		\label{eqn:LN_LC_K_MLE_ARE}
		\mbox{ARE}
		\left( 
		\left(\widehat{\theta}_{\mbox{\tiny K}},
		\widehat{\sigma}_{\mbox{\tiny K}}\right),
		\left(\widehat{\theta}_{\mbox{\tiny MLE}},\widehat{\sigma}_{\mbox{\tiny MLE}}\right)
		\right)
		& =
		\left(
		\mbox{det}
		\left(
		\bm{S}_{\mbox{\tiny MLE}}
		\right)/
		{
			\mbox{det}
			\left(
			\bm{S}_{\mbox{\tiny K}}
			\right)
		}
		\right)^{0.5}
		=
		\left( 
		\dfrac{
			\left(
			c_{2} - c_{1}^{2}
			\right)^{2}
		}
		{
			2
			\left( 
			\Lambda_{1} \Lambda_{3} 
			- 
			\Lambda_{2}^{2}
			\right) 
		}
		\right)^{0.5}.
	\end{align}
	
	It is important to note that the ARE 
	given by Eq. \eqref{eqn:LN_LC_K_MLE_ARE},
	does not depend on the parameters to be estimated. 
	This is not the case for the 
	payment-per-payment loss variable,
	as demonstrated by \cite{MR4263275}
	and \cite{MR4712558}, 
	even for MTM/MWM, 
	where the weights-generating function  
	is much simpler than the one defined in
	Eq. \eqref{eqn:jFun1}.
	
	The numerical values of these AREs, calculated using Eq.~\eqref{eqn:LN_LC_K_MLE_ARE}, are provided in Table~\ref{eqn:LN_LC_K_MLE_ARE}, with the corresponding interaction plot presented in Figure~\ref{fig:LN_ARE} for various combinations of the parameters \( a \) and \( b \). Once again, the maximum efficiency is achieved when both \( a \) and \( b \) are close to 1. Unlike Figure~\ref{fig:ParetoARE}, all ARE values decline sharply for \( a > 1 \), irrespective of the value of \( b \). However, for \( a > 15 \), the resulting ARE interaction curves reach a plateau. This behavior occurs because, for larger values of \( a \) (regardless of \( b \)), \( J(a, b) \) heavily weights higher-order statistics (right-skewed weighting), making these weighting schemes unsuitable for the lognormal model.
	
	\subsection{Fr{\'e}chet Severity Model} 
	
	In this subsection we will choose a severity model based on the extreme value distributions. Extreme value distributions are obtained as limiting distributions of greatest (or least)
	values in random samples of increasing size. Let $X_1,  \cdots, X_n$ be  \textit{iid} with distribution function $F$ and let $M_n= \max\{X_1, X_2,\cdots, X_n\}$.  Assume that there exit sequence  of constants $\{a_n>0\} $ and $\{b_n\}$ such that 
	$\mathbb{P}\left(a_n^{-1}(M_n-b_n)\leq x\right)\rightarrow G(x)$, where $G(x)$ is a non-degenerate cdf. Then according to the  Fisher-Tippett-Gnedenko Theorem 
	\cite{ft28}, 
	the limiting distribution $G(x)$ is the cdf of a Gumbel, a Fr{\'e}chet or  a Weibull distribution. For example, the maximum order statistic from a Pareto distribution converges to a Fr\'{e}chet distribution as the sample size approaches infinity.
	
	The Fr\'{e}chet (Fr)  distribution 
	(also known as the inverse Weibull distribution) 
	is a special case of the generalized distribution of extreme values. 
	This type II   extreme value distribution is used
	to model maximum values in a dataset,
	such as flood analysis, maximum rainfall,
	survival analysis, among others \cite{MR1892574}. 
	As a member of the location-scale family 
	of distributions, the pdf and cdf are,
	respectively, given by 
	\begin{eqnarray*}
		f(x)
		&=& 
		\dfrac{\alpha}{\sigma}
		\left( 
		\dfrac{x - \theta}{\sigma}
		\right)^{-(\alpha + 1)}
		\exp 
		\left[ 
		- 
		\left( 
		\dfrac{\sigma}{x - \theta}
		\right)^{\alpha}
		\right]; 
		\quad 
		x > \theta, \\[5pt]
		F(x) 
		&=&
		\exp 
		\left[
		-\left( 
		\frac{\sigma }{x - \theta}
		\right)^{^{\alpha }}
		\right],
	\end{eqnarray*}
	where 
	$-\infty < \theta < \infty$, 
	$\sigma > 0$, and $\alpha > 0$
	are, respectively, 
	the location, scale, and shape parameters. 
	Since our focus is on estimating 
	financial claim severity models, 
	we set the location parameter 
	$\theta = 0$.
	Thus, 
	we are left to estimate the scale parameter, $\sigma$, 
	and the shape parameter, $\alpha$,
	of the distribution. 
	The moments and quantile functions
	are given by
	\begin{eqnarray*}
		\E 
		\left[ 
		X^k
		\right] 
		&=&
		\sigma^{k} \,
		\Gamma
		\left(
		1-\frac{k}{\alpha}
		\right);
		\quad 
		k < \alpha, \\
		F^{-1}(u) 
		& = & 
		\sigma 
		\left( - \log (u) \right)^{-1/\alpha}.
	\end{eqnarray*}
	
	Like in lognormal model, 
	we take the two functions
	\begin{align}
		\label{eqn:FHFun1}
		h_{1}(x)
		& = 
		\log{(x)}
		\quad 
		\mbox{and} 
		\quad 
		h_{2}(x) 
		= 
		\left( 
		\log{(x)}
		\right)^{2}.
	\end{align}
	
	Thus, with $H_{j} := h_{j} \circ F^{-1}$,
	it follows that 
	\begin{eqnarray}
		& & 
		H_{1}(u) 
		=
		h_{1}
		\left(
		F^{-1}(u)
		\right) 
		= 
		\log{(\sigma)} 
		- 
		\dfrac{\log{(-\log{(u)})}}{\alpha}, 
		\nonumber \\
		\implies 
		& & 
		H_{1}^{'}(u) 
		= 
		- 
		\dfrac{1}{\alpha u \log{(u)}},
		\label{eqn:FDerH1} \\ 
		& & 
		H_{2}(u) 
		=
		h_{2}
		\left(
		F^{-1}(u)
		\right) 
		= 
		\left( 
		\log{(\sigma)}
		\right)^{2}
		- 
		\dfrac{2 \log{(\sigma)} \log{( -\log{(u)})}}{\alpha} 
		+ 
		\dfrac{1}{\alpha^{2}} 
		\left( 
		\log{( -\log{(u)})}
		\right)^{2}, 
		\nonumber \\
		\implies & & 
		H_{2}^{'}(u) 
		= 
		\frac{2 \log{( -\log(u))}}{\alpha^{2} u \log(u)}
		-
		\frac{2 \log(\sigma)}{\alpha u \log(u)}
		=
		\frac{2}{\alpha u \log{(u)}}
		\left( 
		\alpha^{-1} 
		\log{( -\log(u))}
		- 
		\log{(\sigma)}
		\right).
		\label{eqn:FDerH2}
	\end{eqnarray}
	
	Eqs. \eqref{eqn:S1} and \eqref{eqn:P1}
	respectively take the following form
	\begin{eqnarray}
		& & 
		\begin{cases}
			\widehat{\mu}_{1}
			=
			\dfrac{1}{n}
			\displaystyle 
			\sum_{i=1}^{n}
			J \left( \dfrac{i}{n+1} \right) 
			\log{(X_{i:n})}, \\[15pt] 
			\widehat{\mu}_{2}
			=
			\dfrac{1}{n}
			\displaystyle 
			\sum_{i=1}^{n}
			J \left( \dfrac{i}{n+1} \right) 
			\left(\log{(X_{i:n})}\right)^{2},
		\end{cases}
		\label{eqn:FS1} \\[15pt]
		& & 
		\begin{cases}
			\mu_{1}
			=
			\int_{0}^{1}
			J(u)H_{1}(u) \, du 
			= 
			\log{(\sigma)} 
			- 
			\dfrac{1}{\alpha} \kappa_{1}, \\[15pt]
			\mu_{2}
			=
			\int_{0}^{1}
			J(u)H_{2}(u) \, du 
			= 
			\left(\log{(\sigma)}\right)^{2}
			- 
			\dfrac{2 \log{(\sigma)} \kappa_{1}}{\alpha} 
			+ 
			\dfrac{\kappa_{2}}{\alpha^{2}},
		\end{cases}
		\label{eqn:FP1}
	\end{eqnarray}
	where the two integrals 
	$\kappa_{1}$ and $\kappa_{2}$
	are given by 
	\begin{align}
		\label{eqn:IntDefn1}
		\kappa_{1} 
		& 
		\equiv 
		\kappa_{1}(J) 
		: = 
		\int_{0}^{1}
		J(u) 
		\log{\left( - \log{(u)} \right)} \, du 
		\mbox{ and }  
		\kappa_{2} 
		\equiv 
		\kappa_{2}(J) 
		: = 
		\int_{0}^{1}
		J(u) 
		\left( 
		\log{\left( - \log{(u)} \right)}
		\right)^{2} \, du 
	\end{align}
	do not depend on the parameters 
	$\sigma$ and $\alpha$ to be estimated. 
	
	Setting 
	$
	\left(
	\mu_{1}, \mu_{2} 
	\right) 
	= 
	\left( 
	\widehat{\mu}_{1}, \widehat{\mu}_{2}
	\right)
	$
	from Eqs.~\eqref{eqn:FS1} and 
	\eqref{eqn:FP1}, 
	and solving for $\sigma$ and $\alpha$, 
	we get the explicit $L$-estimators as 
	\begin{eqnarray}
		\label{eqn:FEst1}
		\begin{cases}
			\widehat{\alpha}_{\mbox{\tiny K}}
			=
			\sqrt{
				\dfrac{\tau}
				{\widehat{\mu}_{2} - \widehat{\mu}_{1}^{2}}
			}
			=:
			g_{1}
			\left( 
			\widehat{\mu}_{1}, 
			\widehat{\mu}_{2}
			\right), 
			\quad 
			\mbox{where} 
			\quad 
			\tau 
			\equiv 
			\tau(J) 
			:= 
			\kappa_{2} - \kappa_{1}^{2}, \\[10pt]
			\widehat{\sigma}_{\mbox{\tiny K}}
			= 
			\exp 
			\left\{ 
			\widehat{\mu}_{1}
			+ 
			\dfrac{\kappa_{1}}
			{\widehat{\alpha}_{\mbox{\tiny K}}}
			\right\}
			=:
			g_{2}
			\left( 
			\widehat{\mu}_{1}, 
			\widehat{\mu}_{2}
			\right).
		\end{cases}
	\end{eqnarray}
	
	\medskip 
	
	\begin{note}
		Even though the Fréchet distribution is a member
		of the location-scale family and includes 
		a shape parameter, 
		we do not consider the \( h \)-functions 
		as defined in Eq.~\eqref{eqn:LS_hFun_Def1}, 
		since they do not lead to explicit formulas
		for the estimated parameters,
		as shown in Eq.~\eqref{eqn:FEst1}, 
		which results from a different choice 
		of \( h \)-functions from Eq.~\eqref{eqn:FHFun1}.
		That is, 
		\( F_0^{-1}(u) = \left( -\log(u) \right)^{-1/\alpha} \) 
		is not independent of the shape parameter, 
		as \( \alpha \) remains present in the expression. 
		\qed 
	\end{note}
	
	\begin{note}
		From Corollary \ref{cor:IntRelation1}(i), it follows that the estimated 
		\( \widehat{\alpha}_{\mbox{\tiny K}} \) in Eq.~\eqref{eqn:FEst1} is a positive real number (as expected), provided that 
		\(
		\widehat{\mu}_{2} - \widehat{\mu}_{1}^{2} > 0,
		\)
		which may not hold for certain weights-generating functions \( J(\cdot) \), as discussed in Note \ref{note:M2MinusMu1Neg1}. In such cases, the weights-generating function \( J(\cdot) \) must be adjusted to ensure that 
		\(
		\widehat{\mu}_{2} - \widehat{\mu}_{1}^{2} > 0.
		\)
		\qed
	\end{note}
	
	Based on
	Eqs.~\eqref{eqn:FDerH1} and 
	\eqref{eqn:FDerH2}, 
	the entries for the matrix 
	$\bm{\Sigma}$ from Theorem \ref{thm:CGJ1} 
	are given by
	\begin{eqnarray}
		\sigma_{11}^{2} 
		& = & 
		\int_{0}^{1} 
		\int_{0}^{1} 
		J(v) J(w)
		K(v,w) \, 
		dH_{1}(v) \, 
		dH_{1}(w) 
		\nonumber \\
		& = &
		\dfrac{1}{\alpha^{2}}
		\int_{0}^{1} 
		\int_{0}^{1} 
		\dfrac{J(v) J(w) K(v,w)}
		{v w \log{(v)} \log{(w)}}
		dv \, dw 
		\nonumber \\
		& = & 
		\dfrac{1}
		{\alpha^{2}}
		\Psi_{1}(a,b), \\
		\sigma_{12}^{2}
		& = &
		\int_{0}^{1} 
		\int_{0}^{1} 
		J(v) J(w)
		K(v,w) \, 
		dH_{1}(v) \, 
		dH_{2}(w) 
		\nonumber \\
		& = &
		\frac{2}
		{\alpha^{2}} 
		\int_{0}^{1} 
		\int_{0}^{1}  
		\frac{J(v) J(w)
			K(v,w)}{v w \log{(v)} \log{(w)}
		}
		\left( 
		\log{(\sigma)}
		-
		\alpha^{-1} 
		\log{( -\log(w))}
		\right)
		dv \, dw 
		\nonumber \\
		& = & 
		\frac{2 \log{(\sigma)} }
		{\alpha^{2}} 
		\Psi_{1}(a,b)
		- 
		\dfrac{2}{\alpha^{3}} 
		\int_{0}^{1} 
		\int_{0}^{1}  
		\frac{J(v) J(w) K(v,w) \log{( -\log(w))}}
		{v w \log{(v)} \log{(w)}
		}
		dv \, dw
		\nonumber \\
		& = & 
		\frac{2 \log{(\sigma)}}
		{\alpha^{2}} 
		\Psi_{1}(a,b)
		-
		\frac{2}
		{\alpha^{3}}
		\Psi_{2}(a,b), \\
		\sigma_{22}^{2}
		& = &
		\int_{0}^{1} 
		\int_{0}^{1} 
		J(v) J(w)
		K(v,w) \, 
		dH_{2}(v) \, 
		dH_{2}(w) 
		\nonumber \\
		& = &
		\dfrac{4}{\alpha^{2}}
		\int_{0}^{1} 
		\int_{0}^{1} 
		\dfrac{ J(v) J(w) K(v,w)} 
		{{v w \log{(v)} \log{(w)}}}
		\left( 
		\frac{\log{( -\log(v))}}{\alpha}
		- 
		\log{(\sigma)}
		\right)
		\left( 
		\frac{\log{( -\log(w))}}{\alpha}
		- 
		\log{(\sigma)}
		\right)
		dv \, dw 
		\nonumber \\ 
		& = & 
		\dfrac{4 \left( \log{(\sigma)} \right)^{2}}
		{\alpha^{2}}
		\int_{0}^{1} 
		\int_{0}^{1} 
		\dfrac{ J(v) J(w) K(v,w)} 
		{{v w \log{(v)} \log{(w)}}} 
		dv \, dw 
		\nonumber \\ 
		& & 
		- 
		\dfrac{4 \log{(\sigma)}}{\alpha^{3}}
		\int_{0}^{1} 
		\int_{0}^{1} 
		\dfrac{ J(v) J(w) K(v,w) 
			\log{( -\log(v))}} 
		{{v w \log{(v)} \log{(w)}}} 
		dv \, dw 
		\nonumber \\ 
		& & 
		- 
		\dfrac{4 \log{(\sigma)}}{\alpha^{3}}
		\int_{0}^{1} 
		\int_{0}^{1} 
		\dfrac{ J(v) J(w) K(v,w) 
			\log{( -\log(w))}} 
		{{v w \log{(v)} \log{(w)}}} 
		dv \, dw 
		\nonumber \\ 
		& &
		+ 
		\dfrac{4}{\alpha^{4}}
		\int_{0}^{1} 
		\int_{0}^{1} 
		\dfrac{ J(v) J(w) K(v,w) 
			\log{( -\log(v))} \log{( -\log(w))}} 
		{{v w \log{(v)} \log{(w)}}} 
		dv \, dw 
		\nonumber \\ 
		& = & 
		\dfrac{4 \left( \log{(\sigma)} \right)^{2}}
		{\alpha^{2}}
		\Psi_{1}(a,b)
		- 
		\dfrac{8 \log{(\sigma)}}{\alpha^{3}}
		\Psi_{2}(a,b)
		+ 
		\dfrac{4}{\alpha^{4}}
		\Psi_{3},
	\end{eqnarray}
	with the obvious double integration
	notations for 
	$\Psi_{i} \equiv \Psi_{i}(a,b)$ 
	for $i = 1, 2, 3$.
	
	\begin{cor}
		\label{cor:IntRelation1}
		For all \( a > 0 \) and \( b > 0 \), 
		the following inequalities hold: 
		\begin{itemize}
			\item[(i)]
			\( 
			\tau 
			=
			\kappa_{2}
			-
			\kappa_{1}^{2}
			>
			0.
			\)
			
			\item[(ii)] 
			\( 
			\Psi_{1} \, \Psi_{2} 
			- 
			\Psi_{2}^{2}
			> 
			0.
			\)
		\end{itemize}
		
		\begin{proof}
			As in Corollary \ref{cor:LSCov1},
			we again use Theorem \ref{thm:GeneralCSIneq1}
			and Lemma \ref{lemma:KS_TF1} to establish 
			the required inequalities. 
			\begin{itemize} 
				\item[(i)]
				This immediately follows by taking 
				\( 
				g(x)
				= 
				\log{ 
					\left( 
					-
					\log{(x)}
					\right)
				}
				\) 
				in Lemma \ref{lemma:KS_TF1}.
				
				\item[(ii)]
				This follows from Theorem \ref{thm:GeneralCSIneq1}
				with the following assignments:
				\begin{align*}
					f_{1}(x) 
					& =  
					\dfrac{J(x)}
					{
						x \, \log{(x)} 
					}
					\quad 
					\mbox{and}
					\quad 
					f_{2}(x) 
					= 
					\log{ 
						\left( 
						-
						\log{(x)}
						\right)
					}.
					\qedhere 
				\end{align*}
			\end{itemize}
		\end{proof}
	\end{cor}
	
	The matrix $\bm{D}$ is now calculated as 
	\begin{eqnarray}
		\label{eqn:FmD1}
		{\bm D} 
		& = & 
		\left[d_{ij}\right]_{2\times 2}
		=
		\left[\left. \frac{\partial g_{i}}{\partial \widehat{\mu}_{j}}\right\vert_{\widehat{\bm{\mu}}
			=
			\bm{\mu}}\right]_{2 \times 2} 
		\nonumber \\[5pt]
		& = & 
		\begin{bmatrix}
			\dfrac
			{\mu_{1} 
				\sqrt{\tau}
			}
			{
				\left( \mu_{2} 
				- 
				\mu_{1}^{2} 
				\right)^{3/2}}
			& 
			- 
			\dfrac
			{ 
				\sqrt{\tau}
			}
			{
				2
				\left( \mu_{2} 
				- 
				\mu_{1}^{2} 
				\right)^{3/2}}
			\\[20pt] 
			\sigma
			\left( 
			1 
			- 
			\dfrac{\mu_{1} \kappa_{1}}
			{\sqrt{
					\tau
					\left( 
					\mu_{2} - \mu_{1}^{2} 
					\right)
				}
			}
			\right) & 
			\dfrac{
				\sigma 
				\kappa_{1} 
			}
			{2
				\sqrt{
					\tau
					\left( 
					\mu_{2} - \mu_{1}^{2} 
					\right)
				}
			}
		\end{bmatrix} 
		\nonumber \\[10pt]
		& = & 
		\begin{bmatrix}
			\dfrac{
				\alpha^{2}
				\left( 
				\alpha \log{(\sigma)} - \kappa_{1}
				\right) 
			}
			{\tau}
			& 
			- 
			\dfrac{
				\alpha^{3}
			}
			{2 \tau }
			\\[20pt] 
			\dfrac{
				\sigma 
				\left( 
				\kappa_{2} 
				- 
				\alpha \log{(\sigma)} \kappa_{1}
				\right) 
			}
			{\tau}
			& 
			\dfrac{
				\alpha \sigma \kappa_{1}
			}
			{2 
				\tau 
			}
		\end{bmatrix}.
	\end{eqnarray}
	
	Then, it follows that 
	\begin{align}
		\label{eqn:FAsym1}
		\left( 
		\widehat{\alpha}_{\mbox{\tiny K}}, 
		\widehat{\sigma}_{\mbox{\tiny K}}
		\right) 
		& \sim  
		\mathcal{AN}
		\left( 
		\left(\theta, \sigma \right), 
		\dfrac{1}{n} \bm{S}_{\mbox{\tiny K}}
		\right), 
	\end{align}
	where 
	\begin{align}
		\label{eqn:FAsym2}
		\bm{S}_{\mbox{\tiny K}}
		& = 
		\bm{D} \bm{\Sigma} \bm{D}' 
		\nonumber \\
		& = 
		\dfrac{1}{\tau^{2}}
		\begin{bmatrix}
			\alpha^{2} 
			\left(
			\Psi_{1} \kappa_{1}^{2} 
			- 
			2 \Psi_{2} \kappa_{1} 
			+
			\Psi_{3}
			\right) & 
			\sigma 
			\left(
			\Psi_{2} \kappa_{2} 
			-
			\Psi_{3} \kappa_{1} 
			+ 
			\Psi_{2} \kappa_{1}^{2} 
			- 
			\Psi_{1} \kappa_{1} \kappa_{2}
			\right) \\[5pt]
			\sigma 
			\left(
			\Psi_{2} \kappa_{2} 
			-
			\Psi_{3} \kappa_{1} 
			+ 
			\Psi_{2} \kappa_{1}^{2} 
			- 
			\Psi_{1} \kappa_{1} \kappa_{2}
			\right) & 
			\alpha^{-2} 
			\sigma^{2}
			\left(
			\Psi_{3} \kappa_{1}^{2} 
			- 
			2 \Psi_{2} \kappa_{1} \kappa_{2}
			+ 
			\Psi_{1} \kappa_{2}^{2}
			\right)
		\end{bmatrix}.
	\end{align}
	
	From Eq.~\eqref{eqn:FAsym2},
	it immediately follows that
	\begin{align}
		\label{eqn:FAsym3}
		\mbox{det}
		\left( 
		\bm{S}_{\mbox{\tiny K}}
		\right) 
		& = 
		\dfrac{\sigma^{2}
			\left( 
			\Psi_{1} \Psi_{3} 
			- 
			\Psi_{2}^{2}
			\right)}
		{\tau^{2}}
		> 
		0, 
		\quad 
		\mbox{from Corollary \ref{cor:IntRelation1}.}
	\end{align}
	
	\begin{table}[b!]
		\caption{  
			$
			\mbox{ARE}
			\left( 
			\left(\widehat{\theta}_{\mbox{\tiny K}},
			\widehat{\sigma}_{\mbox{\tiny K}}\right),
			\left(\widehat{\theta}_{\mbox{\tiny MLE}},\widehat{\sigma}_{\mbox{\tiny MLE}}\right)
			\right)
			$
			for selected values of $a$ and $b$.
		}
		\label{table:FrechetARE}
		\centering
		\begin{tabular}{|c|cccccccccc|}
			\hline 
			\multirow{2}{*}{$a$} & 
			\multicolumn{10}{|c|}{$b$} \\ 
			\cline{2-11} 
			{} & 0.3 & 0.5 & 0.8 & 1 & 1.3 & 2 & 5 & 7 & 15 & 20 \\
			\hline 
			0.3  & 0.041 & 0.190 & 0.483 & 0.472 & 0.372 & 0.206 & 0.073 & 0.065 & 0.073 & 0.083 \\
			0.5  & 0.032 & 0.163 & 0.674 & 0.845 & 0.824 & 0.575 & 0.172 & 0.108 & 0.043 & 0.034 \\
			0.8  & 0.024 & 0.115 & 0.536 & 0.795 & 0.965 & 0.953 & 0.536 & 0.399 & 0.191 & 0.142 \\
			1.0  & 0.021 & 0.096 & 0.451 & 0.691 & 0.882 & 0.962 & 0.681 & 0.555 & 0.324 & 0.260 \\
			1.2  & 0.019 & 0.083 & 0.386 & 0.603 & 0.794 & 0.917 & 0.748 & 0.643 & 0.427 & 0.360 \\
			2.0  & 0.014 & 0.055 & 0.244 & 0.393 & 0.544 & 0.696 & 0.733 & 0.697 & 0.586 & 0.542 \\
			4.0  & 0.009 & 0.031 & 0.127 & 0.207 & 0.296 & 0.405 & 0.508 & 0.518 & 0.513 & 0.505 \\
			5.0  & 0.008 & 0.026 & 0.103 & 0.167 & 0.241 & 0.334 & 0.432 & 0.447 & 0.457 & 0.455 \\
			7.0  & 0.007 & 0.020 & 0.074 & 0.121 & 0.175 & 0.246 & 0.331 & 0.347 & 0.368 & 0.371 \\
			10.0 & 0.005 & 0.015 & 0.053 & 0.085 & 0.124 & 0.176 & 0.244 & 0.259 & 0.281 & 0.287 \\
			\hline 
		\end{tabular}
	\end{table}
	
	From \cite{MR3706798} we also note that 
	the maximum likelihood estimator of 
	$\sigma$ is given by 
	\begin{align}
		\widehat{\sigma}_{\mbox{\tiny MLE}}
		& = 
		\left(
		\dfrac{1}{n} 
		\sum_{i=1}^{n} 
		x_{i}^{-\alpha}
		\right)^{-1/\alpha_{\mbox{\tiny MLE}}},
	\end{align}
	where $\alpha_{\mbox{\tiny MLE}}$ is 
	given by the unique zero of the 
	strictly decreasing function
	\[
	\xi(\alpha) 
	=
	\alpha^{-1} 
	+ 
	\sum_{i=1}^{n} x_{i}^{-\alpha}
	\log{(x_{i})}
	\left(
	\sum_{i=1}^{n} x_{i}^{-\alpha}
	\right)^{-1}
	-
	n^{-1} 
	\sum_{i=1}^{n}
	\log{(x_{i})}.
	\]
	
	\begin{figure}[t!]
		\centering
		\includegraphics[width=0.75\textwidth,height=0.50\textheight]{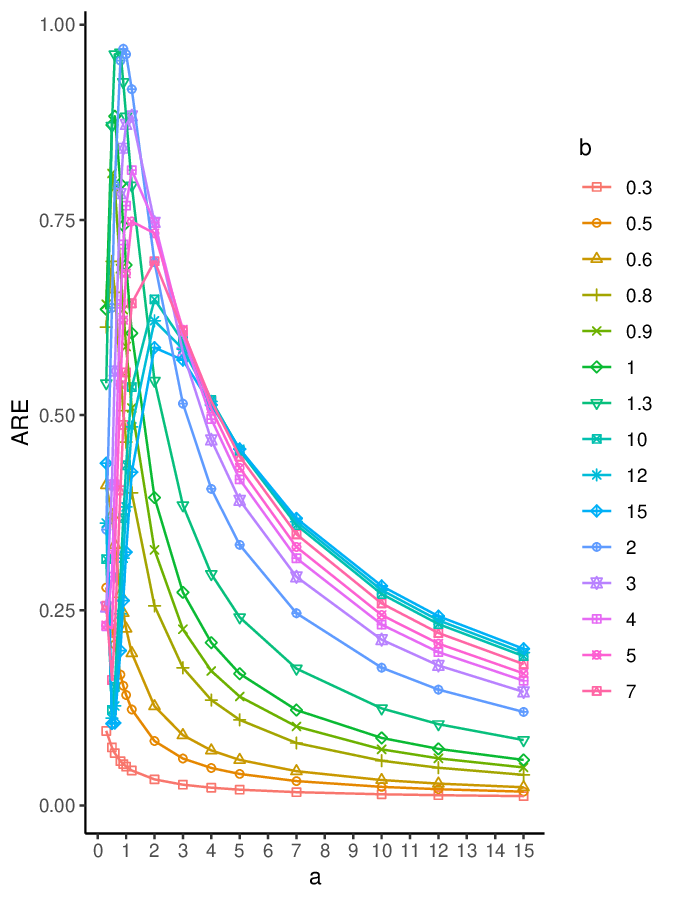} 
		\caption{ARE values of the Fr{\'e}chet severity 
			model presented as interactions between 
			the shape parameters \( a \) and \( b \).}
		\label{fig:FrechetARE}
	\end{figure}
	
	The asymptotic distribution of the MLE, 
	\citet[see, e.g.,][Lemma B.3]{MR3706798}, 
	estimated values is given by
	\begin{align}
		\label{eqn:FThetaAsym1}
		\left( 
		\widehat{\alpha}_{\mbox{\tiny MLE}}, 
		\widehat{\sigma}_{\mbox{\tiny MLE}}
		\right) 
		& \sim  
		\mathcal{AN}
		\left( 
		\left(\alpha, \sigma \right), 
		\dfrac{1}{n} 
		\bm{S}_{\mbox{\tiny MLE}}
		\right),
	\end{align}
	where 
	\begin{align}
		\label{eqn:FThetaAsym2}
		\bm{S}_{\mbox{\tiny MLE}}
		& = 
		\dfrac{6}{\pi^{2}} 
		\begin{bmatrix}
			\alpha^{2} & 
			(\gamma - 1) \sigma \\ 
			(\gamma - 1) \sigma & 
			(\sigma/\alpha)^{2} 
			\left( 
			(\gamma - 1)^{2} + \pi^{2}/6
			\right) 
		\end{bmatrix}
		\quad 
		\mbox{giving}
		\quad 
		\mbox{det}
		\left( 
		\bm{S}_{\mbox{\tiny MLE}}
		\right) 
		=
		\dfrac{6 \sigma^{2}}{\pi^{2}},
	\end{align}
	and 
	$\gamma := - \Gamma'(1) = 0.57721566490$
	is the Euler–Mascheroni constant,
	\citet[][p. 914]{MR3307944}.
	
	Finally, 
	from 
	Eqs.~\eqref{eq:ARE1},
	\eqref{eqn:FAsym3}, and 
	\eqref{eqn:FThetaAsym2},
	it follows that  
	\begin{align}
		\label{eqn:F_K_MLE_ARE}
		\mbox{ARE}
		\left( 
		\left(\widehat{\theta}_{\mbox{\tiny K}},
		\widehat{\sigma}_{\mbox{\tiny K}}\right),
		\left(\widehat{\theta}_{\mbox{\tiny MLE}},\widehat{\sigma}_{\mbox{\tiny MLE}}\right)
		\right)
		& =
		\left(
		{\mbox{det}\left( \bm{S}_{\mbox{\tiny MLE}} \right)}/
		{\mbox{det}
			\left(
			\bm{S}_{\mbox{\tiny K}}
			\right)}
		\right)^{0.5}
		=
		\left( 
		\dfrac{
			6 
			\pi^{-2}
			\tau^{2}
		}
		{
			\Psi_{1} \Psi_{3} 
			- 
			\Psi_{2}^{2}
		}
		\right)^{0.5},
	\end{align}
	which,
	like in Eq.~\eqref{eqn:LN_LC_K_MLE_ARE}, 
	does not depend on the parameters,
	$\alpha$ and $\sigma$, to be estimated. 
	
	The numerical values of these AREs, calculated using Eq.~\eqref{eqn:F_K_MLE_ARE}, are provided in Table~\ref{table:FrechetARE}, with the corresponding interaction plot presented in Figure~\ref{fig:FrechetARE} for various combinations of the parameters \( a \) and \( b \). Similar to the Pareto (Table~\ref{table:ParetoARE}) and lognormal (Table~\ref{table:LNARE}) cases, the maximum efficiency is achieved when both \( a \) and \( b \) are close to 1. The interaction curves in Figure~\ref{fig:ParetoARE} and Figure~\ref{fig:FrechetARE} exhibit almost similar patterns: some curves decrease as \( a \) increases, while others initially increase and then begin to decrease. In contrast, the interaction curves in Figure~\ref{fig:LN_ARE} all follow the same pattern, first increasing and then decreasing. This behavior may reflect the resemblance of the lognormal density pattern to the normal density, as the logarithmic transformation of lognormal random variables results in a normal distribution.
	One plausible explanation for the similarity in behavior observed in Figure~\ref{fig:ParetoARE} and Figure~\ref{fig:FrechetARE} is that both Pareto and Fr{\'e}chet models are characterized by heavy right tails, making them fundamentally distinct from the lognormal distribution,
	which has a relatively lighter tail.
	
	\section{Simulation Study}
	\label{sec:SimStudy}
	
	\begin{figure}[b!]
		\centering
		\includegraphics[width=0.95\linewidth]
		{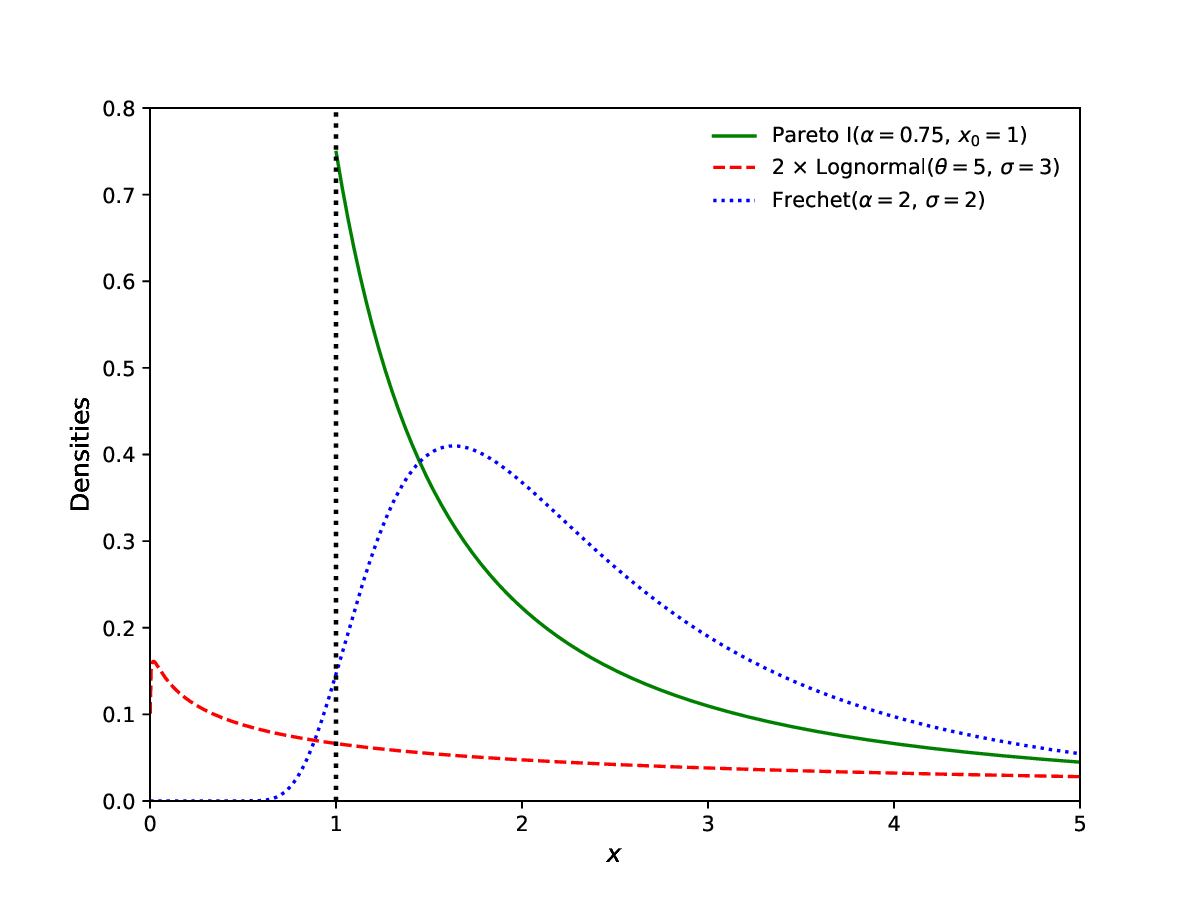}
		\caption{Probability density functions of the 
			three distributions considered in the simulation. 
			For visualization purposes, 
			the pdf of the lognormal distribution 
			is scaled by a factor of 2,
			as indicated in the legend.}
		\label{fig:SimDensities}
	\end{figure}
	
	In this section, 
	we extend the theoretical framework presented in
	Section \ref{sec:ParametricModels}
	by conducting simulations to assess key 
	performance metrics of our estimators for Pareto, Lognormal and Fr\'{e}chet severity models. Probability density functions of the 
	three distributions considered in the simulation study are displayed in Fig.    \ref{fig:SimDensities}.   For visualization purposes, 
	the pdf of the lognormal distribution 
	is scaled by a factor of 2,
	as indicated in the legend.
	
	Specifically, 
	our objectives are to determine the sample size required 
	for estimators to exhibit negligible bias 
	(considering they are asymptotically unbiased),
	to validate asymptotic normality, 
	and to evaluate finite sample relative efficiencies (REs)
	as they approach asymptotic relative efficiencies (AREs).
	For this study, we use the maximum likelihood estimator (MLE) 
	as the baseline for evaluating the relative efficiency of MTM estimators. Therefore, the ARE definition given in Eq.~\eqref{eq:ARE1}
	translates to finite sample performance as follows:
	\begin{equation}
		\label{eqn:FRE1}
		RE(\mathcal{C},\text{MLE})
		= 
		\frac{\text{Asymptotic Variance of MLE}}
		{\text{Small-Sample Variance of $\mathcal{C}$}},
	\end{equation}
	where the numerator corresponds to the
	asymptotic variance given in 
	Eq.~\eqref{eq:ARE1}, 
	and the denominator represents the small-sample 
	variance of the $L$-estimator, expressed as:
	\[
	\left(
	\mbox{det}
	\left\{ 
	\left[ 
	\E 
	\left[ 
	\left( 
	\widehat{\theta}_{i} 
	- 
	\theta_{i}
	\right)
	\left( 
	\widehat{\theta}_{j} 
	- 
	\theta_{j}
	\right)
	\right]
	\right]_{i,j=1}^{k}
	\right\} 
	\right)^{1/2},
	\]
	for a model with \( k \) parameters, 
	\( \bm{\theta} = \left( \theta_{1}, \ldots, \theta_{k} \right) \).
	The simulation design incorporates the elements
	as summarized in Table \ref{table:SimElements1}.
	
	\begin{table}[hbt!]
		\caption{Elements of simulation study.}
		\label{table:SimElements1}
		\centering
		\begin{tabular}{|l|l|l|l|}
			\cline{2-4} 
			\multicolumn{1}{c|}{} & 
			\multicolumn{3}{|c|}{Distribution} \\
			\hline 
			Element & Pareto I & Lognormal & Fr{\'e}chet \\
			\hline\hline 
			Parameter(s) & 
			$\alpha = 0.75$ and $x_{0} = 1$ & 
			$\theta = 5$ and $\sigma = 3$ & 
			$\alpha = 2$ and $\sigma = 2$ \\ 
			\hline 
			Sample Size $(n)$ & 
			$100, \, 250, \, 500, \, 750, \, 1000$ & 
			$100, \, 500, \, 1000$ & 
			$100, \, 500, \, 1000$ \\
			\hline 
			Estimators & 
			\multicolumn{3}{c|}{MLE and $L$-estimators} \\
			\hline 
			$(a,b)$-pairs & 
			\multicolumn{3}{|c|}{
				$(1,1), \, (.8,.8), \, 
				(.8,2), \, (1.2,1.3), \, 
				(4,15), \, (2,.8), \, 
				\mbox{and} \, 
				(5,5)$} \\
			\hline  
		\end{tabular}
	\end{table}
	
	\begin{table}[hbt!]
		\caption{Standardized {\sc Mean} and RE 
			from Pareto I$(\alpha = 0.75, x_{0} = 1)$.
			The entries are mean values 
			(with standard errors in parentheses)
			based on 10,000 samples.}
		\label{table:ParetoSim1}
		\centering
		\begin{tabular}{c|cc|cccccc}
			\cline{2-9} 
			{} & 
			\multicolumn{2}{c|}{KS Par} 
			& 
			\multicolumn{6}{|c}{Sample Size} \\ 
			\cline{2-9} 
			{} & $a$ & $b$ & 100 & 250 & 500 & 750 & 1000 & $\infty$ \\
			\hline\hline 
			\multirow{8}{*}{\rotatebox{90}
				{\sc Mean = $\widehat{\alpha}/\alpha$}} & 
			\multicolumn{2}{|c|}{MLE} & 
			1.01{\tiny (.000)} & 
			1.00{\tiny (.000)} & 
			1.00{\tiny (.000)} & 
			1.00{\tiny (.000)} & 
			1.00{\tiny (.000)} & 
			1 \\
			{} & 1 & 1 & 
			1.01{\tiny (.000)} & 
			1.00{\tiny (.000)} & 
			1.00{\tiny (.000)} & 
			1.00{\tiny (.000)} & 
			1.00{\tiny (.000)} & 
			1 \\
			{} & .8 & .8 & 
			1.05{\tiny (.000)} & 
			1.00{\tiny (.000)} & 
			1.02{\tiny (.000)} & 
			1.01{\tiny (.000)} & 
			1.01{\tiny (.000)} &
			1 \\
			{} & .8 & 2 & 
			1.00{\tiny (.000)} & 
			1.03{\tiny (.000)} & 
			1.00{\tiny (.000)} & 
			1.00{\tiny (.000)} & 
			1.00{\tiny (.000)} &
			1 \\
			{} & 1.2 & 1.3 & 
			0.99{\tiny (.000)} & 
			1.00{\tiny (.000)} & 
			1.00{\tiny (.000)} & 
			1.00{\tiny (.000)} & 
			1.00{\tiny (.000)} & 
			1 \\
			{} & 4 & 15 & 
			1.00{\tiny (.010)} & 
			1.00{\tiny (.000)} & 
			1.00{\tiny (.000)} & 
			1.00{\tiny (.000)} & 
			1.00{\tiny (.000)} & 
			1 \\
			{} & 2 & .8 & 
			1.07{\tiny (.000)} & 
			1.04{\tiny (.000)} & 
			1.02{\tiny (.000)} & 
			1.02{\tiny (.000)} & 
			1.01{\tiny (.000)} & 
			1 \\
			{} & 5 & 5 & 
			0.99{\tiny (.000)} & 
			1.00{\tiny (.000)} & 
			1.00{\tiny (.000)} & 
			1.00{\tiny (.000)} & 
			1.00{\tiny (.000)} & 
			1 \\
			\hline\hline
			\multirow{8}{*}{\rotatebox{90}
				{\sc RE}} & 
			\multicolumn{2}{|c|}{MLE} & 
			0.94{\tiny (.060)} & 
			0.97{\tiny (.050)} & 
			1.00{\tiny (.060)} & 
			1.00{\tiny (.040)} & 
			1.00{\tiny (.050)} & 1 \\
			{} & 1 & 1 & 
			0.94{\tiny (.063)} & 
			0.97{\tiny (.052)} & 
			1.00{\tiny (.061)} & 
			1.00{\tiny (.038)} & 
			1.00{\tiny (.048)} & 1 \\
			{} & .8 & .8 & 
			0.70{\tiny (.048)} & 
			0.78{\tiny (.036)} & 
			0.83{\tiny (.051)} & 
			0.84{\tiny (.033)} & 
			0.88{\tiny (.047)} & 0.953 \\
			{} & .8 & 2 & 
			0.69{\tiny (.042)} & 
			0.68{\tiny (.038)} & 
			0.69{\tiny (.033)} & 
			0.69{\tiny (.009)} & 
			0.70{\tiny (.033)} & 0.693 \\
			{} & 1.2 & 1.3 & 
			0.96{\tiny (.058)} & 
			0.95{\tiny (.054)} & 
			0.97{\tiny (.054)} & 
			0.96{\tiny (.014)} & 
			0.98{\tiny (.048)} & 0.964 \\
			{} & 4 & 15 & 
			0.58{\tiny (.037)} & 
			0.58{\tiny (.029)} & 
			0.59{\tiny (.030)} & 
			0.60{\tiny (.023)} & 
			0.59{\tiny (.026)} & 0.596 \\
			{} & 2 & .8 & 
			0.55{\tiny (.033)} & 
			0.64{\tiny (.029)} & 
			0.69{\tiny (.039)} & 
			0.71{\tiny (.028)} & 
			0.75{\tiny (.039)} & 0.856 \\
			{} & 5 & 5 & 
			0.82{\tiny (.051)} & 
			0.81{\tiny (.044)} & 
			0.82{\tiny (.039)} & 
			0.82{\tiny (.036)} & 
			0.82{\tiny (.054)} & 0.820 \\
			\hline 
		\end{tabular}
	\end{table}
	
	We have generated $10^{3}$ samples of a specified 
	size $n$ from each chosen distribution $F$ 
	(specifically, 
	the single-parameter Pareto, 
	lognormal, and 
	Fréchet distributions, 
	with their respective densities
	shown in Figure \ref{fig:SimDensities}
	using Monte Carlo simulations.
	For each sample,
	we estimate the parameters of 
	$F$ with MLE and various $L$-estimators and calculate
	the average mean and RE for these $10^{3}$ estimates. 
	This process is repeated $10$ times,
	and the $10$ resulting means and REs are averaged again,
	with their standard deviations also reported. 
	Repeating this procedure allows us to assess 
	the standard errors of the estimated means and REs, 
	providing findings based on a total of $10^{4}$ samples.
	The reported standardized mean is calculated as the average 
	of $10^{4}$ estimates divided by the true parameter value, 
	with the standard error similarly standardized.
	
	
	\begin{table}[hbt!]
		\caption{Standardized {\sc Mean} and RE 
			from Lognormal$(\theta = 5, \sigma = 3))$.
			The entries are mean values 
			(with standard errors in parentheses)
			based on 10,000 samples.}
		\label{table:LNSim1}
		\centering
		\begin{tabular}{c|cc|cc|cc|cc|cc}
			\cline{2-11}
			{} & 
			\multicolumn{2}{c|}{KS Par} & 
			\multicolumn{2}{|c|}{$n=100$} & 
			\multicolumn{2}{|c|}{$n=500$} &
			\multicolumn{2}{|c|}{$n=1000$} &
			\multicolumn{2}{|c}{$n \to \infty$} \\
			\cline{2-11} 
			& & & & & & & & & & \\[-2.50ex]
			{} & 
			$a$ & $b$ & 
			$\widehat{\theta}/\theta$ & 
			$\widehat{\sigma}/\sigma$ &
			$\widehat{\theta}/\theta$ & 
			$\widehat{\sigma}/\sigma$ &
			$\widehat{\theta}/\theta$ & 
			$\widehat{\sigma}/\sigma$ &
			$\widehat{\theta}/\theta$ & 
			$\widehat{\sigma}/\sigma$ \\
			\hline\hline 
			\multicolumn{1}{c|}{} &
			\multicolumn{10}{c}{} \\[-2.50ex]
			\multirow{8}{*}{\rotatebox{90}{\sc Mean Values}} &
			\multicolumn{2}{|c|}{MLE} &
			1.00{\tiny (.002)} & 0.99{\tiny (.001)}  & 
			1.00{\tiny (.001)} & 1.00{\tiny (.001)} & 
			1.00{\tiny (.001)} & 1.00{\tiny (.000)} & 
			1 & 1 \\
			{} & 
			1 & 1 &
			1.00{\tiny (.002)} & 0.99{\tiny (.001)} & 
			1.00{\tiny (.001)} & 1.00{\tiny (.001)} &
			1.00{\tiny (.001)} & 1.00{\tiny (.000)} &
			1 & 1 \\
			{} & 
			.8 & .8 &
			0.99{\tiny (.002)} & 0.97{\tiny (.001)} &
			1.00{\tiny (.001)} & 0.99{\tiny (.001)} &
			1.00{\tiny (.001)} & 0.99{\tiny (.000)} &
			1 & 1 \\
			{} & 
			.8 & 2 &
			1.00{\tiny (.003)} & 0.96{\tiny (.002)} &
			1.00{\tiny (.001)} & 0.98{\tiny (.002)} &
			1.00{\tiny (.001)} & 0.99{\tiny (.001)} &
			1 & 1 \\
			{} & 
			1.2 & 1.3 &
			1.01{\tiny (.002)} & 1.00{\tiny (.002)} &
			1.00{\tiny (.001)} & 1.00{\tiny (.001)} &
			1.00{\tiny (.001)} & 1.00{\tiny (.001)} &
			1 & 1 \\
			{} &
			4 & 15 & 
			1.00{\tiny (.003)} & 0.91{\tiny (.004)} & 
			1.00{\tiny (.001)} & 0.98{\tiny (.002)} &
			1.00{\tiny (.001)} & 0.99{\tiny (.001)} & 
			1 & 1 \\
			{} & 
			2 & .8 &
			0.95{\tiny (.002)} & 1.01{\tiny (.003)} & 
			0.98{\tiny (.001)} & 1.00{\tiny (.001)} &
			1.00{\tiny (.001)} & 1.00{\tiny (.001)} &
			1 & 1  \\
			{} & 
			5 & 5 &
			1.05{\tiny (.003)} & 0.88{\tiny (.004)} & 
			1.01{\tiny (.001)} & 0.98{\tiny (.002)} &
			1.00{\tiny (.001)} & 0.99{\tiny (.001)} &
			1 & 1 \\
			\hline\hline 
			\multicolumn{1}{c|}{} &
			\multicolumn{9}{c}{} \\[-2.50ex]
			\multirow{8}{*}{\rotatebox{90}{\sc Re Values}} & 
			\multicolumn{2}{|c|}{MLE} &
			\multicolumn{2}{|c}{1.01{\tiny (.021)}} & 
			\multicolumn{2}{|c}{1.00{\tiny (.026)}} &
			\multicolumn{2}{|c}{1.00{\tiny (.029)}} &
			\multicolumn{2}{|c}{1} \\
			{} & 
			1 & 1 &
			\multicolumn{2}{|c|}{1.01{\tiny (.021)}} & 
			\multicolumn{2}{|c|}{1.00{\tiny (.026)}} &
			\multicolumn{2}{|c|}{1.00{\tiny (.029)}} &
			\multicolumn{2}{|c}{1} \\
			{} & 
			.8 & .8 &
			\multicolumn{2}{|c|}{0.95{\tiny (.021)}} & 
			\multicolumn{2}{|c|}{0.92{\tiny (.024)}} &
			\multicolumn{2}{|c|}{0.92{\tiny (.026)}} &
			\multicolumn{2}{|c}{.962} \\
			{} & 
			.8 & 2 &
			\multicolumn{2}{|c|}{0.69{\tiny (.019)}} & 
			\multicolumn{2}{|c|}{0.63{\tiny (.015)}} &
			\multicolumn{2}{|c|}{0.63{\tiny (.018)}} &
			\multicolumn{2}{|c}{.623} \\
			{} & 
			1.2 & 1.3 &
			\multicolumn{2}{|c|}{0.95{\tiny (.020)}} & 
			\multicolumn{2}{|c|}{0.96{\tiny (.024)}} &
			\multicolumn{2}{|c|}{0.97{\tiny (.017)}} &
			\multicolumn{2}{|c}{.974} \\
			{} &
			4 & 15 & 
			\multicolumn{2}{|c|}{0.38{\tiny (.007)}} & 
			\multicolumn{2}{|c|}{0.48{\tiny (.014)}} &
			\multicolumn{2}{|c|}{0.51{\tiny (.017)}} & 
			\multicolumn{2}{|c}{.520} \\
			{} & 
			2 & .8 &
			\multicolumn{2}{|c}{0.62{\tiny (.022)}} & 
			\multicolumn{2}{|c|}{0.63{\tiny (.023)}} &
			\multicolumn{2}{|c|}{0.64{\tiny (.022)}} &
			\multicolumn{2}{|c}{.662} \\
			{} & 
			5 & 5 &
			\multicolumn{2}{|c}{0.34{\tiny (.009)}} & 
			\multicolumn{2}{|c|}{0.49{\tiny (.010)}} &
			\multicolumn{2}{|c|}{0.52{\tiny (.013)}} &
			\multicolumn{2}{|c}{.555} \\
			\hline 
		\end{tabular} 
	\end{table}
	
	
	\begin{table}[hbt!]
		\caption{Standardized {\sc Mean} and RE 
			from Fr{\'e}chet$(\alpha = 2, \sigma = 2)$.
			The entries are mean values 
			(with standard errors in parentheses)
			based on 10,000 samples.}
		\label{table:FrechetSim1}
		\centering
		\begin{tabular}{c|cc|cc|cc|cc|cc}
			\cline{2-11}
			{} & 
			\multicolumn{2}{c|}{KS Par} & 
			\multicolumn{2}{|c|}{$n=100$} & 
			\multicolumn{2}{|c|}{$n=500$} &
			\multicolumn{2}{|c|}{$n=1000$} &
			\multicolumn{2}{|c}{$n \to \infty$} \\
			\cline{2-11} 
			& & & & & & & & & & \\[-2.50ex]
			{} & 
			$a$ & $b$ & 
			$\widehat{\alpha}/\alpha$ & 
			$\widehat{\sigma}/\sigma$ &
			$\widehat{\alpha}/\alpha$ & 
			$\widehat{\sigma}/\sigma$ &
			$\widehat{\alpha}/\alpha$ & 
			$\widehat{\sigma}/\sigma$ &
			$\widehat{\alpha}/\alpha$ & 
			$\widehat{\sigma}/\sigma$ \\
			\hline\hline 
			\multicolumn{1}{c|}{} &
			\multicolumn{10}{c}{} \\[-2.50ex]
			\multirow{8}{*}{\rotatebox{90}{\sc Mean Values}} &
			\multicolumn{2}{|c|}{MLE} &
			1.01{\tiny (.002)} & 1.00{\tiny (.002)}  & 
			1.00{\tiny (.001)} & 1.00{\tiny (.001)} & 
			1.00{\tiny (.001)} & 1.00{\tiny (.000)} & 
			1 & 1 \\
			{} & 
			1 & 1 &
			1.02{\tiny (.002)} & 1.00{\tiny (.002)} & 
			1.00{\tiny (.001)} & 1.00{\tiny (.001)} &
			1.00{\tiny (.001)} & 1.00{\tiny (.000)} &
			1 & 1 \\
			{} & 
			.8 & .8 &
			1.07{\tiny (.002)} & 0.99{\tiny (.002)} &
			1.03{\tiny (.002)} & 1.00{\tiny (.001)} &
			1.00{\tiny (.001)} & 1.00{\tiny (.000)} &
			1 & 1 \\
			{} & 
			.8 & 2 &
			1.01{\tiny (.002)} & 1.00{\tiny (.002)} &
			1.00{\tiny (.001)} & 0.98{\tiny (.001)} &
			1.00{\tiny (.001)} & 1.00{\tiny (.000)} &
			1 & 1 \\
			{} & 
			1.2 & 1.3 &
			1.00{\tiny (.002)} & 1.01{\tiny (.002)} &
			1.00{\tiny (.001)} & 1.00{\tiny (.001)} &
			1.00{\tiny (.001)} & 1.00{\tiny (.000)} &
			1 & 1 \\
			{} &
			4 & 15 & 
			1.12{\tiny (.006)} & 1.02{\tiny (.002)} & 
			1.02{\tiny (.001)} & 1.00{\tiny (.001)} &
			1.01{\tiny (.001)} & 1.00{\tiny (.000)} & 
			1 & 1 \\
			{} & 
			2 & .8 &
			1.09{\tiny (.003)} & 0.99{\tiny (.003)} & 
			1.04{\tiny (.003)} & 1.00{\tiny (.002)} &
			1.02{\tiny (.002)} & 1.00{\tiny (.001)} &
			1 & 1 \\
			{} & 
			5 & 5 &
			1.10{\tiny (.006)} & 1.05{\tiny (.003)} & 
			1.02{\tiny (.002)} & 1.01{\tiny (.001)} &
			1.01{\tiny (.001)} & 1.00{\tiny (.001)} &
			1 & 1 \\
			\hline\hline 
			\multicolumn{1}{c|}{} &
			\multicolumn{9}{c}{} \\[-2.50ex]
			\multirow{8}{*}{\rotatebox{90}{\sc Re Values}} & 
			\multicolumn{2}{|c|}{MLE} &
			\multicolumn{2}{|c}{0.93{\tiny (.041)}} & 
			\multicolumn{2}{|c}{0.99{\tiny (.030)}} &
			\multicolumn{2}{|c}{0.99{\tiny (.031)}} &
			\multicolumn{2}{|c}{1} \\
			{} & 
			1 & 1 &
			\multicolumn{2}{|c|}{0.67{\tiny (.028)}} & 
			\multicolumn{2}{|c|}{0.68{\tiny (.020)}} &
			\multicolumn{2}{|c|}{0.69{\tiny (.020)}} &
			\multicolumn{2}{|c}{.691} \\
			{} & 
			.8 & .8 &
			\multicolumn{2}{|c|}{0.56{\tiny (.024)}} & 
			\multicolumn{2}{|c|}{0.54{\tiny (.015)}} &
			\multicolumn{2}{|c|}{0.54{\tiny (.015)}} &
			\multicolumn{2}{|c}{.536} \\
			{} & 
			.8 & 2 &
			\multicolumn{2}{|c|}{0.94{\tiny (.039)}} & 
			\multicolumn{2}{|c|}{0.95{\tiny (.027)}} &
			\multicolumn{2}{|c}{0.94{\tiny (.025)}} &
			\multicolumn{2}{|c}{.953} \\
			{} & 
			1.2 & 1.3 &
			\multicolumn{2}{|c|}{0.73{\tiny (.029)}} & 
			\multicolumn{2}{|c|}{0.78{\tiny (.024)}} &
			\multicolumn{2}{|c|}{0.78{\tiny (.021)}} &
			\multicolumn{2}{|c}{.794} \\
			{} &
			4 & 15 & 
			\multicolumn{2}{|c|}{0.28{\tiny (.011)}} & 
			\multicolumn{2}{|c|}{0.45{\tiny (.011)}} &
			\multicolumn{2}{|c|}{0.48{\tiny (.014)}} & 
			\multicolumn{2}{|c}{.513} \\
			{} & 
			2 & .8 &
			\multicolumn{2}{|c}{0.28{\tiny (.008)}} & 
			\multicolumn{2}{|c|}{0.26{\tiny (.007)}} &
			\multicolumn{2}{|c|}{0.25{\tiny (.010)}} &
			\multicolumn{2}{|c}{.244} \\
			{} & 
			5 & 5 &
			\multicolumn{2}{|c}{0.24{\tiny (.007)}} & 
			\multicolumn{2}{|c|}{0.38{\tiny (.011)}} &
			\multicolumn{2}{|c|}{0.41{\tiny (.009)}} &
			\multicolumn{2}{|c}{.432} \\
			\hline 
		\end{tabular} 
	\end{table}
	
	The simulation results are presented in Tables 
	\ref{table:ParetoSim1}, 
	\ref{table:LNSim1},
	and \ref{table:FrechetSim1}, 
	corresponding to the Pareto I, 
	lognormal, and 
	Fréchet distributions,
	respectively.
	The entries of the last column(s)
	corresponding to $n \to \infty$ 
	in all three tables 
	are included for completeness and are 
	found in Section \ref{sec:ParametricModels},
	not from simulations. 
	We see that the MEAN
	of all Pareto $\alpha$ estimators converges 
	to the parameter 
	$\alpha$ very fast, 
	and the bias practically disappears 
	when $n \ge 500$, 
	except for $J(0.8, 0.8)$- 
	and $J(2, 0.8)$-weighted $L$-estimators,
	which are heavily right-weighted
	and heavy weights assigned at 
	both tails, respectively, 
	as seen in Figure \ref{fig:KumDensities}.
	For lognormal and Fr{\'e}chet models,
	the convergence of the estimated
	scale parameter, $\sigma$,
	is slower than the estimated location parameter, $\theta$.
	Among all three tables, 
	the maximum relative bias, 
	approximately \(-12\%\), 
	is observed for the \( J(5, 5) \)-weighted estimated value of the scale parameter
	when the sample size is \( n = 100 \).
	
	Similarly, 
	the RE’s converge to their asymptotic 
	counterparts slower in all three tables. 
	Interestingly, 
	for some choices of $a$ and $b$, 
	for example, 
	for $J(0.8, 2)$-weighted RE's for 
	lognormal and 
	for $J(0.8, 0.8)$-weighted RE's for 
	Fr{\'e}chet models are converging 
	to their corresponding asymptotic 
	counter parts from above. 
	
	
	\section{Real Data Analysis}
	\label{sec:RealDataAnalysis}
	
	This section assesses the performance of the estimation methods developed in the previous sections by applying them to a real-world dataset. Specifically, we analyze a dataset consisting of 1500 U.S. indemnity losses, which has been extensively studied in the actuarial literature (see, e.g., \cite{MR1988432}, \cite{MR4712558}). Notably, \cite{tcf24} utilized this dataset to fit Gamma and Inverse-Gaussian models via an intercept-only generalized linear model (GLM) using the score-based weighted likelihood estimation approach, highlighting its suitability for robust modeling.
	This dataset has been recognized as fitting the lognormal model. 
	
	As summarized in Table~\ref{table:US_IndemnitySampled1} (top portion), the Kolmogorov-Smirnov (KS) test statistic \citep[see, e.g.,][\S15.4.1, p.~360]{MR3890025} is computed to be 0.0266. With a significance level of 5\%, the corresponding critical value is 0.0351. These results confirm that the lognormal model is a plausible fit for the indemnity loss dataset at the 5\% significance level, thereby reinforcing its relevance for this empirical analysis.
	To further illustrate the benefits of the proposed weighted robust fitting approach, 
	the dataset was slightly modified 
	by replacing its largest observation; 2,173,595, with $10^{7}$ (10 million).
	
	First, we fit the lognormal severity models to the dataset using both MLE and the $J(1.1, 1.2)$-weighted $L$-estimators, where the $J(1.1, 1.2)$-weighted model assigns slightly lighter weights to both tails of the data compared to the observed sample values, resembling the $J(1.2, 1.3)$ curve shown in Figure~\ref{fig:TWKTwoColPy1}. 
	As presented in Table~\ref{table:US_IndemnitySampled1} (top portion), while both MLE and $J(1.1, 1.2)$-weighted models produce similar fitted values for $\theta$ and $\sigma$ across both the original and modified datasets, the $J(1.1, 1.2)$-weighted model demonstrates greater stability. Specifically, for the original dataset, the $J(1.1, 1.2)$-weighted model achieves a $p$-value of $29.02\%$, which is $5.26\%$ higher than the $p$-value of $23.76\%$ obtained from the MLE model.
	
	In contrast, for the modified dataset, the $p$-value of the $J(1.1, 1.2)$-weighted model increases relative to the original dataset, while the $p$-value for the MLE model decreases. This indicates the robustness of the $J(1.1, 1.2)$-weighted model under data perturbations. Additionally, the KS-test statistic values are consistently lower for the $J(1.1, 1.2)$-weighted model compared to the corresponding MLE results, highlighting its superior performance and stability.
	
	Second, 
	purely for illustrative purposes and to enable a clearer visualization, 
	as it is challenging to represent all 1,500 sample observations, a random subsample of size 50 is extracted using \verb|seed(123)|. 
	This subsample is utilized to observe and demonstrate the stability of the proposed $L$-estimation approach. 
	The sampled data are presented below:
	\begin{center}
		\begin{tabular}{*{10}{r}}
			\toprule
			1,000 & 3,436 & 5,000 & 7,500 & 9,000 & 10,899 & 14,500 & 20,000 & 30,000 & 95,000 \\
			1,500 & 3,486 & 5,000 & 7,525 & 9,250 & 11,667 & 15,000 & 25,000 & 30,000 & 153,874 \\
			1,913 & 4,000 & 5,010 & 8,500 & 9,500 & 12,100 & 19,500 & 25,187 & 32,000 & 337,500 \\
			2,500 & 5,000 & 6,000 & 8,939 & 10,000 & 12,875 & 20,000 & 30,000 & 65,000 & 412,998 \\
			2,500 & 5,000 & 6,750 & 9,000 & 10,199 & 14,500 & 20,000 & 30,000 & 74,970 & 2,173,595 \\
			\bottomrule
		\end{tabular}
	\end{center}
	
\begin{figure}[hbt!]
		\centering
		\includegraphics[width=1.00\textwidth]{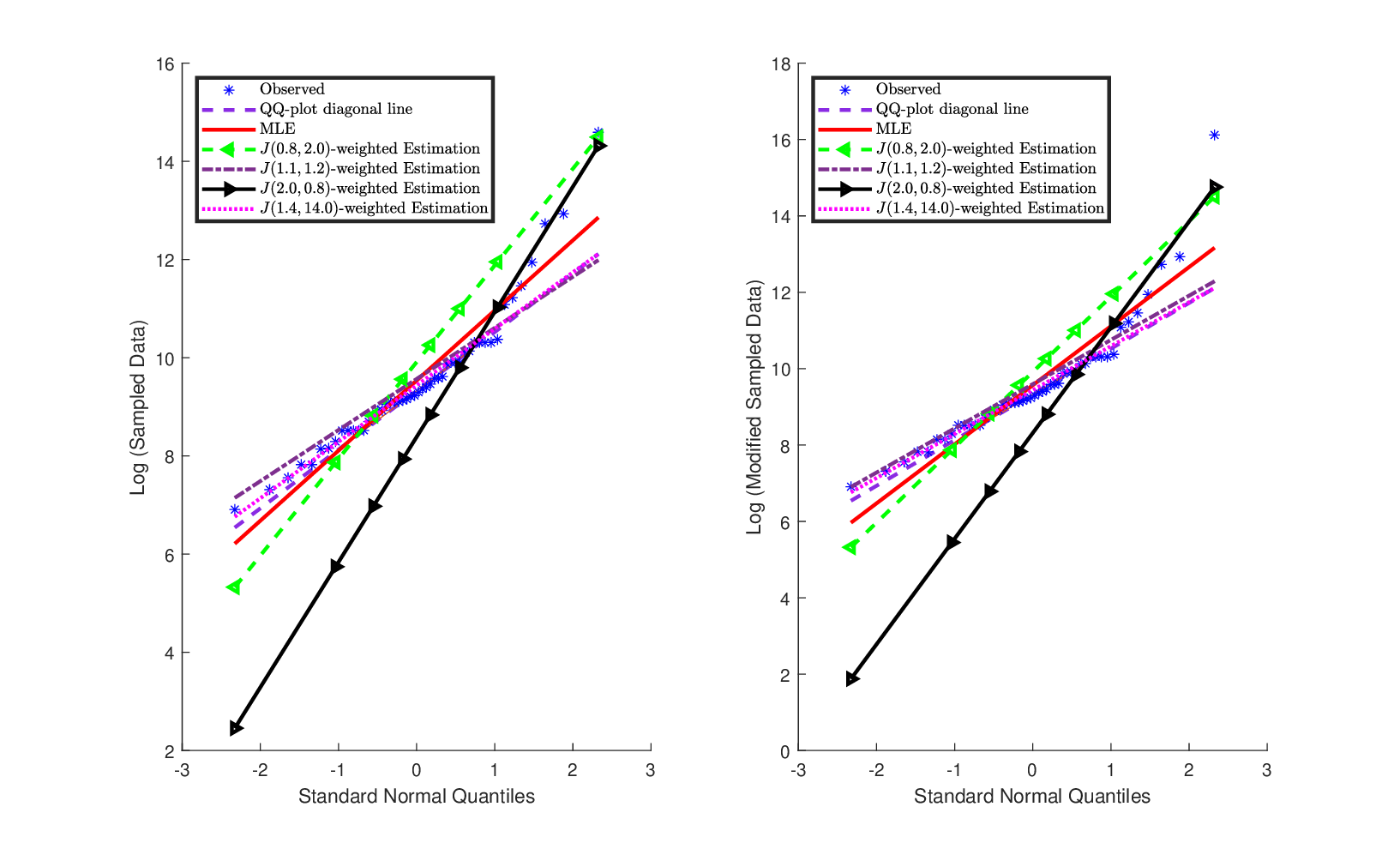}
		\caption{Quantile-quantile plot and
			lognormal fits for the sampled US Indemnity loss data (left panel) and 
			the modified version of the 
			sampled data (right panel).}
		\label{fig:KS_Indemnity}
	\end{figure}

	Similar to the approach applied to the original dataset, 
	we modified the sampled dataset to demonstrate the advantages
	of weighted robust fitting. 
	Specifically,
	the largest observation, 2,173,595, 
	was replaced with $10^{7}$ (10 million). 
	We then refit the lognormal models to the sampled dataset
	using MLE and various $J(a,b)$-weighted $L$-estimators, 
	as specified in 
	Table~\ref{table:US_IndemnitySampled1} (bottom portion).
	The resulting fits are visualized in 
	Figure~\ref{fig:KS_Indemnity}—left panel 
	for the sampled dataset and right panel 
	for the modified sampled dataset, 
	respectively. 
	The numerical parameter estimates and goodness-of-fit metrics 
	are presented in 
	Table~\ref{table:US_IndemnitySampled1} (bottom portion).
	
	\begin{table}[hbt!]
		\caption{MLE and Kumaraswamy weighted estimators of 
			$\theta$ and $\sigma$ with their corresponding 
			one-sample Kolmogorov-Smirnov test statistics.}
		\label{table:US_IndemnitySampled1}
		\centering
		\begin{tabular}{|c|c|c|c|c|c|c|c|c|c|c|}
			\hline 
			\multirow{2}{*}{Estimators} & 
			\multirow{2}{*}{$\widehat{\theta}$} & 
			\multirow{2}{*}{$\widehat{\sigma}$} & 
			\multicolumn{3}{|c|}{KS Test} & 
			\multirow{2}{*}{$\widehat{\theta}$} & 
			\multirow{2}{*}{$\widehat{\sigma}$} & 
			\multicolumn{3}{|c|}{KS Test} \\
			\cline{4-6} \cline{9-11}
			{} & 
			{} & {} & 
			$h$ & $p$-value & $D$ & 
			{} & {} & 
			$h$ & $p$-value & $D$ \\
			\hline
			{$-$} & 
			\multicolumn{5}{|c|}{Original Data} & 
			\multicolumn{5}{|c|}{Modified Original Data} \\ 
			\hline\hline  
			MLE & 
			9.374 & 1.638 & 
			0 & 0.2376 & 0.0266 &
			9.375 & 1.641 & 
			0 & 0.2303 & 0.0268 \\ 
			$J(1.1, 1.2)$ & 
			9.381 & 1.627 & 
			0 & \fbox{0.2902} & 0.0252 &
			9.382 & 1.628 & 
			0 & \fbox{0.2932} & 0.0252 \\
			\hline\hline 
			{$-$} &  
			\multicolumn{5}{|c|}{Sampled Data} & 
			\multicolumn{5}{|c|}{Modified Sampled Data} \\ 
			\hline 
			MLE & 
			9.536 & 1.428 & 
			0 & 0.2657 & 0.1387 &
			9.566 & 1.547 & 
			0 & 0.1343 & 0.1609 \\
			$J(0.8, 2.0)$ & 
			9.911 & 1.970 & 
			1 & 0.0012 & 0.2671 &
			9.914 & 1.973 & 
			1 & 0.0012 & 0.2680 \\ 
			$J(1.1, 1.2)$ & 
			9.572 & 0.743 & 
			0 & \fbox{0.3622} & 0.1272 &
			9.595 & 0.887 & 
			0 & 0.3132 & 0.1328 \\ 
			$J(2.0, 0.8)$ & 
			8.386 & 2.549 & 
			1 & 0.0000 & 0.3619 & 
			8.316 & 2.768 & 
			1 & 0.0000 & 0.3749 \\
			$J(1.4, 14.0)$ & 
			9.439 & 1.151 & 
			0 & \fbox{0.8912} & 0.0788 &  
			9.439 & 1.151 & 
			0 & \fbox{0.8912} & 0.0788 \\
			\hline 
		\end{tabular} \\[5pt]
		{\sc Note:}
		{\scriptsize 
			For the KS Test column, \( h \in \{0, 1\} \) 
			represents the hypothesis test result, 
			where \( h = 0 \) indicates that the assumed model is plausible, 
			and \( h = 1 \) indicates that the model is rejected.
			\( D \) denotes the Kolmogorov-Smirnov test statistic, \\
			and the \( p \)-value represents the probability 
			of observing a test statistic at least as extreme 
			as \( D \) under the null hypothesis.
		}
	\end{table}
	
	As shown in Table~\ref{table:US_IndemnitySampled1},
	the \( J(1.1, 1.2) \)- and \( J(1.4, 14) \)-weighted models 
	exhibit significantly higher \( p \)-values compared to the 
	corresponding MLE-fitted model. 
	Notably, transitioning from the MLE to the 
	\( J(1.4, 14) \)-weighted model results 
	in an increase in the \( p \)-value
	from \( 0.2657 \) to \( 0.8912 \), 
	representing a \( 235.42\% \) rise. 
	Furthermore, when moving from the sampled dataset
	to the modified sampled dataset, 
	the \( p \)-value for the MLE-fitted 
	model dropped substantially,
	nearly halving from \( 26.57\% \) 
	to \( 13.43\% \). In contrast, 
	the \( p \)-values for the \( J(1.1, 1.2) \)- 
	and \( J(1.4, 14) \)-weighted models 
	remained virtually unchanged, 
	further highlighting the robustness 
	and stability of the \( J(a,b) \)-weighted models,
	provided the parameters $a$ and $b$ 
	are appropriately selected.
	
	Figure~\ref{fig:KS_Indemnity} offers several insightful observations. 
	Notably, only a few observations 
	(specifically, 
	seven observed stars
	{\blue\textbf{*}}) 
	in the right tail deviate 
	from the straight-line pattern,
	indicating a heavier tail compared
	to the QQ-plot diagonal line 
	\textcolor[rgb]{0.54,0.17,0.89}{\textbf{- -}}.
	However, 
	the remaining data points closely 
	align with the diagonal line.
	In the left panel, 
	representing the sampled dataset, 
	the \( J(1.1, 1.2) \)- and \( J(1.4, 14) \)-weighted 
	fits are closely aligned with the 
	QQ-plot diagonal line, 
	providing a more accurate representation 
	of the data points along the diagonal 
	line compared to the MLE.
	
	In contrast, examining the right panel, 
	which corresponds to the modified sampled dataset, 
	the $J(1.1, 1.2)$- and $J(1.4, 14)$-weighted 
	fits remain robust and unaffected by the data modification, 
	with the data points aligning almost perfectly 
	along the QQ-plot diagonal line. 
	Conversely, the new MLE fit shows significant deviation, 
	particularly in response to the modified largest value of \(10^{7}\), 
	highlighting its sensitivity to extreme observations.
	
	The remaining two fits, \( J(0.8, 2.0) \)- and \( J(2.0, 0.8) \)-weighted models, do not perform as well as the MLE fit, as also observed in Figure~\ref{fig:LN_ARE}.
	This is because \( J(0.8, 2.0) \) assigns disproportionately 
	heavier weights to lower-order statistics 
	(left-skewed weighting), 
	while \( J(2.0, 0.8) \) heavily 
	weights higher-order statistics 
	(right-skewed weighting), 
	as illustrated in Figures \ref{fig:KumDensities} 
	and \ref{fig:TWKTwoColPy1}. 
	These weighting schemes are not well-suited 
	for the sampled dataset under consideration.
	
	\section{Concluding Remarks}
	\label{sec:Conclusion}
	
	This paper presents a flexible and robust \( L \)-estimation 
	framework weighted by Kumaraswamy densities, 
	addressing the limitations of the method of trimmed moments (MTM)
	and the method of winsorized moments (MWM) 
	in modeling claim severity distributions. 
	By incorporating smoothly varying observation-specific weights, 
	the proposed approach effectively balances robustness and efficiency 
	while preserving valuable information from the dataset.
	Explicit formulations for \( L \)-estimators were developed 
	for key parametric models, including Pareto, lognormal, 
	and Fr{\'e}chet distributions, 
	with inferential justification on asymptotic normality
	and variance-covariance structures. 
	The framework was validated through simulations and 
	a real-world dataset of U.S. indemnity losses, 
	demonstrating superior performance in handling 
	outliers and heavy-tailed distributions 
	for predictive loss severity modeling.
	
	The findings of this study suggest several 
	promising directions for future research. 
	First, 
	while the Kumaraswamy-weighted framework is effective 
	for the models examined, 
	its application to broader classes of claim severity 
	and financial models deserves further exploration, 
	including generalized linear models with similar weighting mechanisms.
	Second, 
	extending this methodology to multivariate settings 
	could address dependencies often present in actuarial datasets. 
	
	Third,
	insurance losses often exhibit distributional characteristics 
	such as multimodality and outlier contamination. 
	A current trend in the literature involves fitting 
	spliced or mixture loss severity distributions 
	\citep[see, e.g.,][]{MR3896968, MR4340261, MR4149559}. 
	However, these approaches, 
	typically based on maximum likelihood estimation,
	may lack stability in cost predictions, 
	particularly under data perturbations or outlier contamination. Investigating the applicability of the weighted \( L \)-estimation 
	framework for fitting spliced models presents
	a compelling future direction. 
	Nevertheless, 
	implementing this approach requires the existence
	of quantile functions, as seen in Eq.~\eqref{eqn:P1}, 
	and the inferential justification for
	weighted \( L \)-estimators in spliced models 
	can be highly challenging, if not infeasible.
	In such cases, 
	an algorithmic approach, 
	such as simulation-based estimation, 
	could provide an alternative pathway 
	\citep[see, e.g.,][]{MR3523956}.
	Moreover, 
	integrating the framework with machine learning techniques, 
	including ensemble models or neural networks, 
	could offer innovative solutions for predictive
	modeling in insurance and risk management.
	
	Finally, practical considerations such as computational 
	efficiency and scalability for large datasets remain 
	essential areas for future development. 
	The flexibility and robustness of the proposed 
	framework make it a versatile and promising tool 
	for advancing the precision and stability 
	of actuarial and financial modeling.
	
	\clearpage 
	
	\baselineskip 4.5mm
	\bibliographystyle{apalike}

	
\end{document}